\documentclass[11pt,a4paper]{article}
\usepackage[utf8]{inputenc}
\usepackage{authblk}
\usepackage{amsmath,amsfonts,amssymb}
\usepackage{mathtools}
\usepackage{enumerate}
\usepackage{float}
\usepackage{bm}
\usepackage{upgreek}
\usepackage{rotating}
\usepackage{multicol}
\usepackage{array,multirow}
\usepackage{authblk}
\usepackage{textcomp}
\usepackage{graphicx}
\usepackage{caption}
\usepackage{algorithm}
\usepackage[noend]{algpseudocode}
\usepackage{url,hyperref}
\usepackage{amsthm}

\usepackage{xcolor}
\usepackage{subfig}
\usepackage{float}
\usepackage{longtable}
\usepackage{placeins}
\usepackage[numbers]{natbib}
\numberwithin{equation}{section}

\newcommand{\eps}{\varepsilon}

\newcommand{\cL}{\mathcal{L}}

\newcommand{\cV}{\mathcal{V}}

\newcommand{\cW}{\mathcal{W}}

\newcommand{\bR}{\mathbb{R}}

\newcommand{\bN}{\mathbb{N}}

\newcommand{\bX}{\mathbb{X}}

\newcommand{\bfx}{\mathbf{x}}

\makeatletter\@addtoreset{equation}{section} \makeatother

 \allowdisplaybreaks

\begin{document}
\title{Generative Pricing of Basket Options via Signature-Conditioned Mixture Density Networks}
\author[1]{Hasib Uddin Molla \thanks{Corresponding author: \url{mdhasibuddin.molla@ucalgary.ca}}}
\author[1]{Antony Ware}
\author[2]{Ilnaz Asadzadeh}
\author[2]{Nelson Mesquita Fernandes}
\affil[1]{Department of Mathematics and Statistics, University of Calgary, Calgary, Canada}
\affil[2]{BMO Capital Markets, Toronto, Canada}

\maketitle

\begin{abstract}
We present a generative framework for pricing European-style basket options by learning the conditional terminal distribution of the log arithmetic-weighted basket return. A Mixture Density Network (MDN) maps time-varying market inputs encoded via truncated path signatures to the full terminal density in a single forward pass. Traditional approaches either impose restrictive assumptions or require costly re-simulation whenever inputs change, limiting real-time use. Trained on Monte Carlo (MC) under GBM with time-varying volatility or local volatility, the MDN acts as a reusable surrogate distribution: once trained, it prices new scenarios by integrating the learned density. Across maturities, correlations, and basket weights, the learned densities closely match MC (low KL) and produce small pricing errors, while enabling \emph{train-once, price-anywhere} reuse at inference-time latency. \\

\end{abstract}

%%%%%%%%%%%%%%%%%%%%%%%%%%%%%%%%%%%%%%%%%%%%%%%%%%% Intro
\section{Introduction}\label{sec1}

Accurate and efficient option pricing is fundamental to risk management and trading strategies in modern financial markets, particularly for institutional investors. Traditional pricing methods, such as the Black–Scholes model or numerical techniques like Monte Carlo simulations, remain widely used but exhibit notable limitations. Closed-form models often rely on restrictive assumptions, while numerical methods can be computationally expensive and struggle to capture complex market features, particularly in multi-asset portfolios or models with local volatility and time-varying parameters.\\

\noindent A promising alternative is to learn the terminal density of basket returns across a range of market inputs and then compute prices using risk-neutral expectations. However, this requires an accurate representation of the conditional probability density function (PDF) of the underlying asset under realistic dynamics—including time-dependent interest rates, dividend yields, local volatility surfaces, and asset correlations. In such cases, a closed-form expression for the probability density function of the underlying distribution is not available. In some cases, a semi-closed-form formula is available but requires additional complex computations. Furthermore, these solutions are typically limited to a single set of model parameters, which restricts their practicality in dynamic or data-driven settings. \\

\noindent To address these limitations, recent advances in deep learning offer a promising avenue. In particular, deep neural networks—when trained across a suitably chosen parameter space—can approximate the underlying asset's density function with high flexibility and generalization capability. This transforms the pricing task into a “single solve” problem, meaning that once the model is trained, it can be used to price options for arbitrary combinations of input parameters, enabling efficient and scalable valuation across various market regimes.

\subsection{From Risk-Neutral Expectation to Density Learning}
Consider an asset price process $S(t;\boldsymbol{\vartheta})$, defined on the time interval $t\in [0,T]$, where $\boldsymbol{\vartheta}$ denotes the set of model parameters. For a given payoff function  $f(S(T;\boldsymbol{\vartheta}),K)$, representing an option with maturity $T$ and strike price $K$, the corresponding option price $P(T,K;\boldsymbol{\vartheta})$ is given by
\begin{equation}
P(T,K;\boldsymbol{\vartheta})= D_T \mathbb{E}\big[ f(S(T;\boldsymbol{\vartheta});K)\big],
\end{equation}
where $D_T=e^{-rT}$ denotes the discount factor under constant interest rate $r$.

If $p(y|\boldsymbol{\vartheta}, t)$ denotes the conditional probability density function of $S(t;\boldsymbol{\vartheta})$, then the expectation can be equivalently expressed as the integral:
\begin{equation}\label{pfhj}
P(T,K;\boldsymbol{\vartheta})=D_T \int f(y;K)p(y|\boldsymbol{\vartheta},T) dy.
\end{equation}

\noindent Hence, once the terminal conditional density $p(y|\boldsymbol{\vartheta},T)$ of the underlying asset is known or accurately approximated, the option price can be computed for arbitrary parameters directly from this integral representation.

\subsection{Mixture Density Networks for Distribution Approximation}
Mixture models provide a flexible framework for approximating complex distributions by representing them as convex combinations of simpler components, typically Gaussian. Such mixtures capture multimodality, skewness, and other nonlinear features observed in empirical return distributions. A Mixture Density Network (MDN) extends this idea by using a neural network to output both the mixture weights and component parameters, allowing the learned model to represent highly flexible conditional distributions.

MDNs have proven effective in modelling stochastic systems across diverse fields, from acoustic modelling \citep{Zen_2014_MDNAcousticModeling} and bioinformatics \citep{Ji_2005_BetaMixtureBioinfo} to probabilistic forecasting \citep{Zhang_2020_RegionalWindMDN} and volatility estimation. Several studies have also explored the financial applications of MDNs. Schittenkopf and Dorffner \citep{Schittenkopf_2001_RND_MDNet} extended the MDN framework to extract risk-neutral densities from observed option prices, while Schittenkopf et al. \citep{Schittenkopf_1998_MDNVolatility} utilized MDNs to estimate and forecast stock market volatility. Li et al. \citep{Li_2024_XRMDN} introduced a recurrent MDN architecture for demand forecasting in high-volatility mobility-on-demand systems. 

Nevertheless, most existing work focuses on static settings, leaving open the challenge of incorporating time-varying and path-dependent features—such as evolving interest rates and dividend yields—into a unified probabilistic pricing framework.

\subsection{Signature-Conditioned MDNs for Basket Option Pricing}

 In this work, we propose a signature-conditioned Mixture Density Network for generative pricing of European-style basket options. The model learns to approximate the conditional terminal distribution of basket returns under realistic market dynamics. These conditions include local volatility, asset correlations, and time-dependent interest rates and dividend yields. 

 A key modelling challenge is encoding time-varying parameters in a format suitable for deep learning. The use of truncated path signatures from rough path theory addresses this by providing a finite-dimensional and scalable feature representation that captures the essential temporal information of parameter trajectories. When combined with the MDN, they allow the model to efficiently and robustly learn mappings from dynamic market trajectories to the terminal distribution of the basket return.

 Once trained, the model produces accurate option prices by integrating the learned density against the payoff function of European call and put options. This positions the MDN as a generative surrogate that replaces re-simulation at inference time. Empirical results demonstrate that our method achieves pricing accuracy comparable to benchmark Monte Carlo simulations while offering substantial computational speedups.\\
 
\noindent \textbf{Relation to Generative Models in Finance: }The Mixture density network approach for conditional density approximation belongs to the broader class of generative models. MDNs directly model $p(y|\textbf{x})$ as a finite mixture and train via exact negative log-likelihood (NLL). In one-dimensional settings, this yields a lightweight, numerically stable, and inherently multi-modal representation. By contrast, variational autoencoders (VAEs) optimize an evidence-lower-bound (ELBO) surrogate \citep{Kingma2014VAE, Rezende2014StochasticBackprop} and introduce additional encoder-decoder structures; while flexible, they add unnecessary latent complexity for 1D conditionals and are prone to posterior collapse. Normalizing flows provide exact likelihoods \citep{Papamakarios2017MAF, Durkan2019NSF, Papamakarios2021NFJMLR}  and excel in high-dimensional outputs, but are often over-parameterized for 1D tasks unless carefully constrained. Finally, diffusion and score-based models \citep{Ho2020DDPM,Song2021SDE} offer state-of-the-art generative fidelity but require multi-step sampling and heavy conditioning, leading to higher computational cost. For the univariate conditional densities considered in this study, the MDN strikes the best balance between fidelity, numerical stability, and deployment cost. Nevertheless, flow- or diffusion-based frameworks remain promising candidates for future multivariate extensions, where richer joint structures among asset components become critical.
 
\subsection{Background on Path Signatures}
The signature of a path, introduced by Chen \citep{chen_1957_IntgofPaths, chen_1977_ItePathIntg}, is defined as an infinite sequence of iterated integrals of the path over increasing tensor orders. In practice, one often works with a truncated signature, and this truncation is theoretically justified by the fact that any path of finite $p$-variation is uniquely determined by its signature truncated at order $\lfloor p \rfloor$ (see Chapter 7 of \citep{Friz_2010_RoughPaths}). Its ability to capture the essential characteristics of time series in a non-parametric manner has led to successful applications in various domains. In financial time series analysis, signatures have been effectively employed for both classification and prediction tasks \citep{GyurkoLyons_2014_Extracting_signature_financial_data, LyonsNi_2014_feature_set_financialdata}, demonstrating their utility in handling complex, high-frequency data. Beyond predictive modelling, signatures have been applied to solve forward-backward stochastic differential equations (FBSDEs) numerically \citep{QiFeng_2023_DeepSignatureFBSDEAlgo}, particularly in non-Markovian settings where path-dependence plays a critical role. \\

\noindent The remainder of this paper is organized as follows.
Section \ref{sec:MDN_path_signatures} reviews the theoretical foundations of Mixture Density Networks and truncated path signatures.
Section \ref{sec:steps_training_MDN} describes the model setup, data generation under Geometric Brownian Motion with local volatility and time-varying rates, and the training procedure.
Section \ref{sec:Numerical_Experiments} presents numerical experiments comparing our approach to benchmark Monte Carlo pricing.
Finally, Section \ref{sec:conclusion} concludes with key insights and directions for future research.

%%%%%%%%%%%%%%%%%%%%%%%%%%%%%%%%%%%%%%%%%

\section{Mixture Density Networks and Path Signatures} \label{sec:MDN_path_signatures}
In this section, we provide a brief overview of the core components of our framework: the mixture density model, the mixture density network (MDN), and path signatures.
\subsection{Mixture Density Model (Univariate)}
A mixture model represents a complex probability distribution as a weighted sum of simpler component distributions. Formally, an arbitrary probability density function $p(y)$ can be approximated by a finite mixture of $d$ component densities $\phi_j(y)$ with corresponding mixing weights $\pi_j$ for $j=1,\cdots,d$:
 \begin{equation}
     p(y)= \sum_{j=1}^d \pi_j \phi_j(y|\boldsymbol{\lambda}_j),
 \end{equation} 
 where $\boldsymbol{\lambda}_j$ denotes the parameters of the $j$-th component distribution, determining its shape, scale and location. When component distribution and mixing coefficients are appropriately selected, this framework can approximate a wide variety of continuous probability distributions with high fidelity. In practice, mixtures of Gaussian distributions are frequently used due to their analytical tractability and universal approximation properties.

For the univariate target distribution $p(y)$, we consider univariate component distributions in the mixture model. For the Gaussian mixture model (GMM) of a univariate distribution, 
$\boldsymbol{\lambda}_j=(\mu_j,\delta_j)$ and 
 \begin{equation}
     \phi_j(y|\mu_j,\delta_j)=\frac{1}{\sqrt{2\pi \delta_j^2}} e^{-\frac{(y-\mu_j)^2}{2\delta_j^2}},
 \end{equation}
where, $\mu_j$ are mean and $\delta_j>0$ are standard deviation of $j$-th Gaussian distribution; and $\pi_j>0$ decides how $d$-Gaussian distributions are mixed together and must satisfy the normalization condition $\sum_{j=1}^d \pi_j =1$.

The flexibility of the mixture model lies in its ability to approximate arbitrary probability density functions by adjusting the means $\mu_j$, standard deviations $\delta_j$, and mixing coefficients $\pi_j$ of the component Gaussians.

\subsection{Mixture Density Network (Maximum Likelihood)}
A mixture density network (MDN) is a hybrid architecture that combines a neural network with a mixture density model to approximate complex conditional probability distributions. Given an input (conditioning) feature vector $\mathbf{x}$, an univariate MDN predicts the parameters of a (Gaussian) mixture model:
\begin{itemize}
\item Mixing coefficients, $\pi(\mathbf{x})=(\pi_1(\mathbf{x}),\cdots,\pi_d(\mathbf{x})),$
\item Means, $\mu(\mathbf{x})=(\mu_1(\mathbf{x}),\cdots,\mu_d(\mathbf{x})),$
\item Standard deviations, $\delta(\mathbf{x})=(\delta_1(\mathbf{x}),\cdots,\delta_d(\mathbf{x})),$
\end{itemize}

and finally the conditional density $p(y|\mathbf{x})$ is given by
\begin{equation}
    p(y|\mathbf{x}) = \sum_{j=1}^d \pi_j(\mathbf{x})\times \phi_j(y|\mu_j(\mathbf{x}),\delta_j(\mathbf{x})).
\end{equation}

The network takes $\mathbf{x}\in\bR^{m_0}$ as input and produces the three parameter vectors $\pi(\mathbf{x}), \mu(\mathbf{x})$ and $\delta(\mathbf{x})$. These outputs are subject to specific constraints to ensure they define a valid Gaussian mixture:
\begin{itemize}
\item Mixing coefficients $\pi_j(\mathbf{x})$ must be non-negative and sum to 1. This is typically enforced by applying the $\textit{softmax}$ function to the raw network outputs for $\pi$.
    \item Standard deviations $\delta_j(\mathbf{x})$ must be strictly positive. A common choice is the exponential function \citep{Bishop_1994_MDN} to map unconstrained outputs to positive values, though alternatives such as $\textit{ELU}$ \citep{Normandin_2023_LinearPretrainRMDN, Wu_2022_3DHumanPoseMDN, Yang_2021_GANMDNInverseModeling} and $\textit{softplus}$ \citep{Goodfellow_2016_DeepLearning, Han_2022_SurvivalMDN, Hepp_2022_MDNIntervals, Muzakka_2024_RBSMDN, Razavi_2024_FRMDN, Wang_2022_LikelihoodFreeMDN} are sometimes preferred to improve numerical stability.
    \item Means $\mu(\mathbf{x})$ are unconstrained real numbers and are typically output directly without transformation.
\end{itemize}

\begin{figure}[]
        \centering
\includegraphics[width=0.8\textwidth]{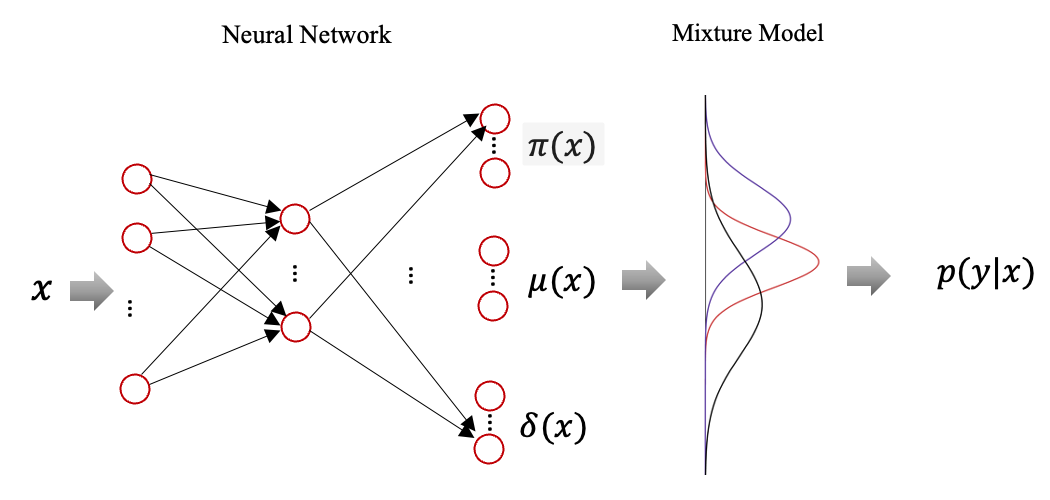}
        \caption{Mixture Density Network}
    \label{fig:MDN_architecture_general}
    \end{figure}

This structure enables the MDN to learn an arbitrary conditional distribution $p(y|\mathbf{x})$ from data. A general schematic of the MDN is shown in Figure \ref{fig:MDN_architecture_general}. \\

\noindent We now describe the internal structure of the MDN. First, the input layer calculates
\begin{equation}
    y_0=g(\cW_0\mathbf{x}+\beta_0 ),
\end{equation}
where $y_0\in \bR^m$; $m$ is the number of neurons in each hidden layer, $\cW_0$ and $\beta_0$ are the weight matrix and bias vector of the input layer, and $g$ is the activation function. 

Subsequently, each hidden layer $h=1,\cdots,H$ computes:
\begin{equation}
    y_h=g(\cW_h y_{h-1}+\beta_h ),
\end{equation}
where $y_h\in\bR^m$; $\cW_h$ and $\beta_h$ are the weight matrices and bias vectors of the $h$-th hidden layer. 

The final layer outputs the parameters of the mixture model as:
\begin{align*}
    \pi &=g_{\pi}(\cW_{\pi} y_H+\beta_{\pi} ),\\
    \mu &=g_{\mu}(\cW_{\mu} y_H+\beta_{\mu}) ,\\
    \delta &=g_{\delta}(\cW_{\delta} y_H+\beta_{\delta} ),
\end{align*}
where $g_{\pi},g_{\mu}$ and $g_{\delta}$ are activation functions chosen to ensure appropriate constraints on the output. \\

The hidden layers of the network apply nonlinear transformations to the input, enabling the MDN to capture complex dependencies between $\mathbf{x}$ and the conditional distribution $p(y|\mathbf{x})$.
The choice of network architecture and activation functions in an MDN is problem-specific and may be adapted to ensure both numerical stability and appropriate modelling behaviour. In our implementation, we further customize the activation functions to (i) impose the initial condition exactly on the PDF and (ii) regularize the density in the presence of singularities.\\

\noindent\textbf{Training Objective (Exact Likelihood).} The network is trained by negative log-likelihood (NLL) with a \emph{LogSumExp} stabilization:

\begin{align}
     \log p(\mathbf{y}^i|\mathbf{x}^i) &= \sum_{k=1}^M\log \left( \sum_{j=1}^d \pi_j(\mathbf{x}^i)\times \phi_j(y^{i,k}|\mu_j(\mathbf{x}^i),\delta_j(\mathbf{x}^i))\right)\nonumber\\
     & = \sum_{k=1}^M\log \left( \sum_{j=1}^d \exp{\Big( \log \pi_j(\mathbf{x}^i)+ \log\big(\phi_j(y^{i,k}|\mu_j(\mathbf{x}^i),\delta_j(\mathbf{x}^i))\big)\Big)}\right),
\end{align}
and 
\begin{align}
    \cL_{\text{NLL}}& = -\frac{1}{N_s} \sum_{i=1}^{N_s} \log p(\mathbf{y}^i|\mathbf{x}^i) .
\end{align}

\subsection{Path Signatures}

In this work, we use the truncated signature of trajectories as features in the MDN. Let $T>0$ be a fixed time horizon, and let $m\in\bN,0<p\in\bR$. Denote by $\cV^p([0,T];\bR^m)$ the space of continuous paths $X:[0,T]\rightarrow \bR^m$ of finite $p$-variation. The signature of the path $X$ over the interval $I=[0,T]$ is defined as the infinite sequence of its iterated integrals:

\begin{equation}\label{sigwhole}
Sig(X)_{0,T}:= \Big(1,\; \bX^1_{0,T},\cdots,\bX^m_{0,T}, \;\bX^{1,1}_{0,T},\; \bX^{1,2}_{0,T}, \cdots, \bX^{m,m}_{0,T},  \;\bX^{1,1,1}_{0,T},  \cdots  \Big),
\end{equation}
where each element corresponds to a multi-indexed iterated integral of the form

\begin{equation}\label{sigdefn}
\bX_{0,T}^{i_1,\cdots, i_l}:=\int_{0<t<T} \bX_{0,t}^{i_1,\cdots, i_{l-1}}dX^{i_l}_t=\int_{0<t_l<T}\cdots \int_{0<t_1<t_2}dX_{t_1}^{i_1}\cdots dX_{t_l}^{i_l},
\end{equation}
for $l\geq 1$ and $i_j\in\{1,\cdots,m\}$.
The zeroth level of the signature is defined to be $1$ by convention. The first level of the signature is the collection of $m$ real numbers $\bX^1_{0,T},\cdots,\bX^m_{0,T}$ and the second level is the collection of $m^2$ real numbers $\bX^{1,1}_{0,T},\cdots,\bX^{1,m}_{0,T},\bX^{2,1}_{0,T},\cdots,\bX^{m,m}_{0,T}$. More generally, the $l$-th level of the signature consists of all terms $\bX^{i_1, \cdots,i_l}_{0,T}$ over the multi-indices of length $l$.

When the path $(X_t)_{t\in[0,T]}$ is of finite variation, the iterated integrals in \eqref{sigdefn} are interpreted as Riemann-Stiltjes integrals. When $(X_t)_{t\in[0,T]}$ is a continuous semi-martingale (e.g., Brownian motion or asset price processes or other stochastic processes), the integrals in \eqref{sigdefn} are in the Stratonovich sense, aligning with rough path theory.

The truncated signature of $X$ up to level $l$ is denoted by $Sig(X)^l_{0,T}$ and is obtained by retaining all terms in the infinite sequence in \eqref{sigwhole} up to (and including) level $l$. This truncation yields a finite-dimensional feature representation that preserves the distinguishable characteristics of the path. One can choose $l$ to balance fidelity vs. parameter count; in our experiments, $l$ around 4-5 works well.
% \subsection{From Density to Prices}
% The MDN outputs $p(y|\mathbf{x})$ for the terminal $y=\log\left(\sum_{j=1}^N w_j\frac{S_j(T)}{S_j(0)}\right)$, where $w_j$ is the relative weight of the $j$th asset in the basket. European prices are then expectations under this learned density (via quadrature or Monte Carlo sampling from the mixture). This positions the MDN as a generative surrogate that replaces re-simulation at inference time.

\section{The Mechanism of Training an MDN}\label{sec:steps_training_MDN}

\begin{figure}[]
    \centering
\includegraphics[width=1.0\textwidth]{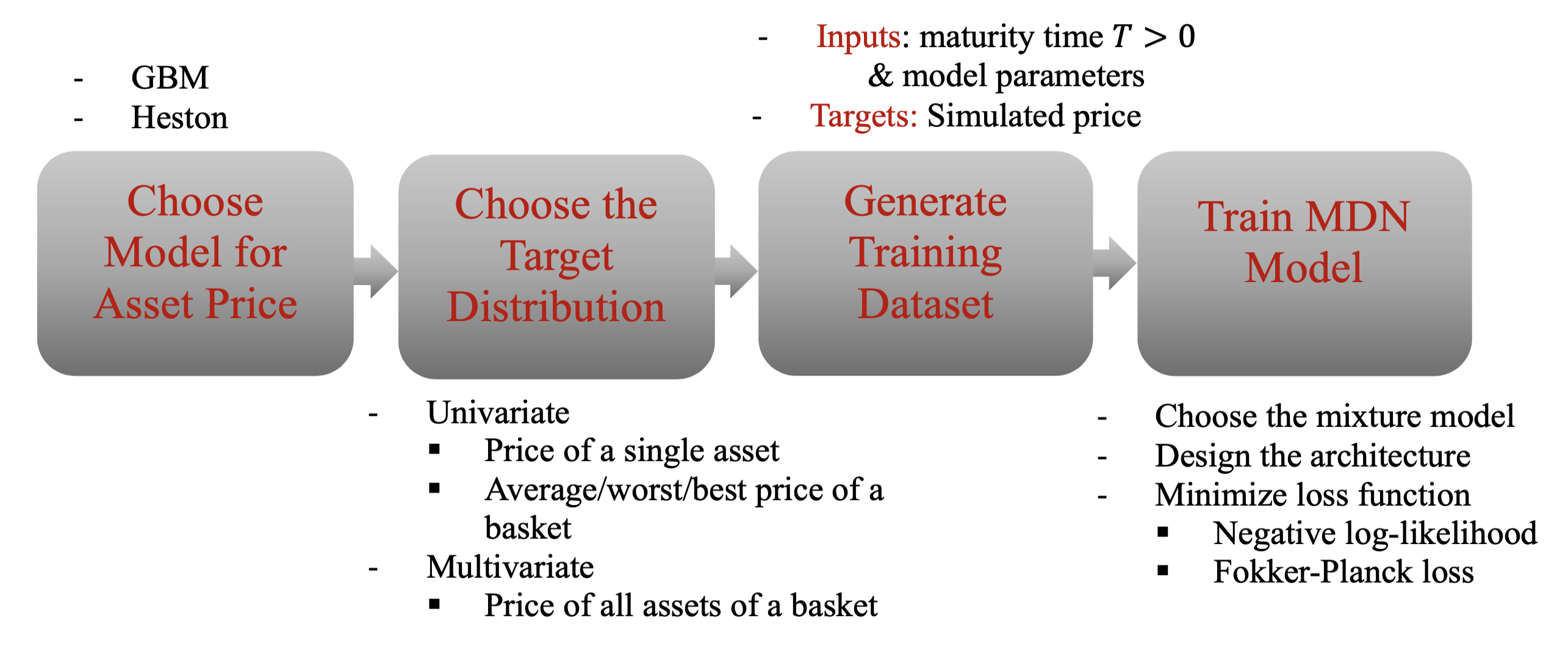}
     \caption{MDN Training Workflow}
    \label{fig:training_workflow_MDN}
\end{figure}

Training a Mixture Density Network (MDN) to price a specific derivative product involves several key decision points. A schematic overview of the MDN training process is presented in  Figure \ref{fig:training_workflow_MDN}. \\

\noindent Next, we will discuss all these aspects in detail.

\subsection{Asset Price Model}
Asset price dynamics in financial markets are commonly modelled using stochastic differential equations (SDEs), with popular choices including the Geometric Brownian Motion (GBM), the Heston stochastic volatility model, and various rough volatility models. In this work, we focus on models in which the asset price evolves under a GBM framework with generalized volatility structures.

\subsubsection{Geometric Brownian Motion (GBM)}
Let $r(t)$ denote the risk-free interest rate and $q_j(t)$ the dividend rate of asset $j$. Consider a portfolio of $N$ assets whose price processes are denoted by $S(t)=(S_1(t),\cdots, S_N(t))$. Each asset $S_j(t)$ follows the SDE:
	\begin{equation}
		dS_j(t)=(r(t)-q_j(t))S_j(t) dt + \sigma_j(t, S_j(t)) S_j(t) dW_j(t),\; S_j(0)=s_{0,j},
	\end{equation}
for  $t\in[0,T]$ and $j=1,2,\cdots,N,$ where $W_j(t)$ denotes the standard Brownian motion, and $\sigma_j(\cdot,\cdot)$ is the volatility function. 
In this work, we consider two forms of volatility for each asset:
\begin{enumerate}[i.]
\item Time dependent volatility: $\sigma_j(t,S_j(t))=\sigma_j(t).$\\
In this setting, volatility varies with time but is independent of the asset’s current price. This can model market regimes where volatility follows a deterministic term structure.
\item Local volatility: $\sigma_j(t,S_j(t))=\sigma_j(S_j(t)).$\\
Here, volatility is a function of the asset’s spot price, allowing the model to capture empirical features such as volatility smiles and skews.
\end{enumerate}
These formulations enable the modelling of more realistic market dynamics than the classical constant-volatility GBM.

\subsubsection{Correlation Structure of the Basket}
In modelling a basket of $N$ assets, the correlation structure among their respective Brownian motions is an essential component. Let $\mathbf{R}\in \bR^{N\times N}$ denote the correlation matrix governing the $N$-dimensional Brownian motion vector:
\begin{equation}
    (W_1, \cdots, W_N)^T.
\end{equation}
Assuming that $\mathbf{R}$ is positive semi-definite and symmetric, it admits a Cholesky decomposition $\mathbf{R}=\mathbf{LL}^T$, where $\mathbf{L}$ is a lower triangular matrix with strictly positive diagonal entries. Then, the correlated Brownian motion increments can be expressed as:
\begin{equation}
    (dW_1(t), \cdots, dW_N(t))^T = \mathbf{L}(d\widetilde{W}_1(t),\cdots, d\widetilde{W}_N(t))^T,
\end{equation}
where $d\widetilde{W}_1(t),\cdots, d\widetilde{W}_N(t)$ are independent Brownian motion increments. This construction ensures that the components of the vector $(W_1, \cdots, W_N)$ follow the desired correlation structure specified by $\mathbf{R}$.

\subsection{Target Distribution}

The structure of the option contract determines the choice of the target distribution. In this study, we consider a European-style option written on the weighted return of a basket of $N$ assets. Accordingly, the target is a univariate conditional distribution of the basket's weighted return at a fixed terminal time.

The basket weights can be treated in two ways. If the basket's composition is fixed, we may hold the weights constant throughout training. However, to enable pricing across a variety of portfolio compositions without retraining, the MDN can be parameterized to include the basket weights as part of the input. This allows the model to generalize over different weighting schemes and provides greater flexibility in practical applications.\\

\noindent \textbf{$\log$-space as the target of MDN:} To improve the accuracy of the MDN approximation, we transform the target variable to log-space. Specifically, the MDN is trained to learn the distribution of the logarithm of the basket's average return, resulting in a smoother, more stable learning target.

\subsection{Training Data Generation}

The generation of training data for the MDN is closely tied to both the choice of the underlying asset price model and the MDN's parameterization. This section is divided into two subsections, each corresponding to a distinct volatility structure assumed under Geometric Brownian Motion (GBM) dynamics for the underlying assets. For each case, we provide a comprehensive list of model parameters along with the complete algorithm used to generate the training data.

The numerical experiments presented later in the paper are also organized according to these two cases. Before detailing the case-specific procedures, we first outline the general sampling strategies employed for the option maturity, the lower-triangular matrix $\mathbf{L}$ obtained from the Cholesky decomposition of the correlation matrix, and the time-varying market rates.\\

\noindent\textbf{Sampling for maturity, $T$:} We define maturity $T$ in the interval $(0,1]$, where $T=1$ corresponds to one year. To ensure adequate coverage of both short- and long-term behaviours, we sample $T$ from $[0.001, 1.05]$, with more samples drawn near the boundaries.\\

\noindent\textbf{Sampling of $\mathbf{L}$:}  To generate a random sample of a valid correlation matrix $\mathbf{R}$, we begin by randomly sampling a sequence of angles:
\begin{equation}
    \mathbf{\Theta}=\Big\{ \alpha_1, \alpha_2, \cdots ,\alpha_{\frac{N(N-1)}{2}} \Big\},\quad \alpha_i\in(0,\pi),
\end{equation}
 and define
 $$\xi_i = \sin(\alpha_i),\; \gamma_i=\cos(\alpha_i).$$
 
 Using these trigonometric components, we construct a lower triangular matrix $\mathbf{L}\in\bR^{N\times N}$ as follows:
\begin{equation}\label{correlation_matrix_sin_cosine}
    \mathbf{L} =\begin{pmatrix}
  1 & 0 & 0 & \cdots & 0\\ 
   \gamma_1& \xi_1 & 0 & \cdots & 0\\
   \gamma_2 &  \xi_2\gamma_3 & \xi_2 \xi_3 &\cdots & 0 \\
  \vdots & \vdots & \vdots & \ddots & \vdots\\
  \gamma_N & \xi_N  \gamma_{N+1} & \xi_N \xi_{N+1} \gamma_{N+2} &\cdots & \xi_N \xi_{N+1}\cdots \xi_{\frac{N(N-1)}{2}}
\end{pmatrix}.
 \end{equation}
 This construction ensures that the resulting matrix  $\mathbf{LL}^T=\mathbf{R}$ is a valid correlation matrix—i.e., symmetric and positive semi-definite—and allows for flexible specification of inter-asset dependencies through parameterized random sampling.\\

 \noindent \textbf{Sampling of time-varying rates:} As time-varying rates such as interest and dividends are parameters of the GBM model and will become the parameters of the MDN through the mixture model, we will need sample paths of the time-varying parameters to simulate the sample paths for the assets in the basket for the training of the MDN. In practice, such sample paths can be generated from historical data using resampling techniques (e.g., bootstrapping). In our experiments, we adopt the Cox-Ingersoll-Ross (CIR) process to simulate these paths:
\begin{equation}\label{CIR_model}
dX(t)=a(b-X(t))dt + c \sqrt X(t) dW(t),\; X(0)=x_0,
\end{equation}
where $a,b$ and $c$ are fixed, stationary parameters. For each asset $j$ in the basket, we generate random initial values $x_{0,j}$ and simulate the corresponding time-varying paths using the CIR model.\\

Now, we will discuss the details of the training data generation for two different volatility structures under GBM. 

\subsubsection{GBM with Time-varying Volatility}
Consider a basket of $N$ assets, whose price processes $S(t)=(S_1(t),\cdots, S_N(t))$ follow geometric Brownian motion (GBM) dynamics with time-varying volatilities. The stochastic differential equation (SDE) for each asset $j=1,\cdots,N$ is given by:
	\begin{equation}
		dS_j(t)=(r(t)-q_j(t))S_j(t) dt + \sigma_j(t) S_j(t) dW_j(t),\; S_j(0)=s_{0,j},
	\end{equation}
where  $t\in[0,T]$. The Brownian motions $(W_1(t),\cdots,W_N(t))$ are correlated and correlation is specified by a correlation matrix $\mathbf{R}^{N\times N}.$\\

\noindent \textbf{Model parameters: } The completed set of model parameters for the GBM system is denoted by 
$$\boldsymbol{\vartheta}=\Big(r(\cdot), \big(q_j(\cdot), \sigma_j(\cdot)\big)_{j=1}^N, \mathbf{L}\Big),$$  
where $r(\cdot)$ is the time-varying risk-free interest rate, $q_j(\cdot)$ and $\sigma_j(\cdot)$ are the time-varying dividend rate and volatility for $j$th asset in the basket. $\mathbf{L}$ is the lower triangular matrix from the Cholesky decomposition of the correlation matrix $\mathbf{R}$ of the basket, that is, $\mathbf{R}=\mathbf{LL}^T$. The price process $S(t)$ is thus parametrized by $\boldsymbol{\vartheta}$ and denoted by $S(t;\boldsymbol{\vartheta})$. \\

\noindent \textbf{Training set generation:} The Algorithm \ref{alg:param_corr_time_variate_rates_tVGBM} below summarizes the procedure used to generate the training data for the mixture density network (MDN), where the target is the weighted return of the basket with fixed basket weights.

\begin{algorithm}
\caption{Training Set Generation for Time-Varying GBM with Correlated Assets}\label{alg:param_corr_time_variate_rates_tVGBM}
\begin{algorithmic}
\State \textbf{Inputs:}  Initial basket asset price $S(0)$, Basket weight $w$
\State \textbf{Output:} Training dataset ${(\bfx^{m,i}, \mathbf{y}^{m,i})}_{i,m}$
\State 
\State \textbf{Step 1:} Sample $n_1$ maturity values $T^i \in [0, T]$
\State \textbf{Step 2:} For each $i = 1, \dots, n_1$:
\For{$m = 1:n_2$}
\State - Sample initial values $r_0^{m,i}$, $q_0^{m,i}$, and $\sigma_0^{m,i}$
\State - Generate time-varying paths $r^{m,i}(t; r_0^{m,i})$, $q^{m,i}(t; q_0^{m,i})$, and $\sigma^{m,i}(t; \sigma_0^{m,i})$ for $t \in [0, T^i]$ using the CIR model
\State - Sample a lower-triangular matrix $\mathbf{L}^{m,i}$ to define the correlation structure
    \State - Form the parameter set:
    $$\boldsymbol{\vartheta}^{m,i}=\Big(r^{m,i}(\cdot), \big(q_j^{m,i}(\cdot), \sigma_j^{m,i}(\cdot)\big)_{j=1}^N, \mathbf{L}^{m,i}\Big) $$
    \State - Simulate $M$ independent paths of asset price, $\Big\{S^{m,i,k}(t;\boldsymbol{\vartheta}^{m,i})\Big\}_{k=1}^M$ over $t\in [0,T^i]$
\EndFor
\State \textbf{Step 3:} Compute MDN targets:
$$\mathbf{y}^{m,i}=\Bigg\{y^{m,i,k} =\log\bigg( \sum_{j=1}^N w_j \frac{S_j^{m,i,k}(T^i;\;\boldsymbol{\vartheta}^{m,i})}{S_j(0)}\bigg)\Bigg\}_{k=1}^M$$
\State \textbf{Step 4:} Define MDN inputs as $ \bfx^{m,i}=(\boldsymbol{\vartheta}^{m,i},T^i)$
\end{algorithmic}
\end{algorithm}

\subsubsection{GBM with Local Volatility}
We consider a basket consisting of $N$ assets whose price processes $S(t)=(S_1(t),\cdots, S_N(t))$, evolve according to geometric Brownian motion (GBM) dynamics with local volatilities. The dynamics of  each asset $j=1,\cdots,N$ are described by the following stochastic differential equation (SDE):
	\begin{equation}
		dS_j(t)=(r(t)-q_j(t))S_j(t) dt + \sigma_L(S_j(t)) S_j(t) dW_j(t),\; S_j(0)=s_{0,j},
	\end{equation}
where  $t\in[0,T]$ and $\sigma_L(\cdot)$ denote the local volatility function. The Brownian motions $(W_1(t),\cdots,W_N(t))$ are correlated, with the correlation structure captured by a correlation matrix $\mathbf{R}^{N\times N}$.\\

\noindent \textbf{Model parameters: } 
Let $\boldsymbol{\vartheta}=\Big(r(\cdot), \big(q_j(\cdot)\big)_{j=1}^N,\mathbf{L}\Big)$, denote the set of model parameters governing the risk-free interest rate, dividend yields, and the Cholesky factor $\mathbf{L}$ of the correlation matrix. Together with the local volatility parameter $\boldsymbol{\nu}$, the full set of parameters for the GBM model is given by $(\boldsymbol{\vartheta},\boldsymbol{\nu})$. Accordingly, the asset price process is written as $S(t;\boldsymbol{\vartheta},\boldsymbol{\nu})$ to reflect the dependence on both parameter sets.\\

\noindent \textbf{Training set generations:} The procedure for generating the training dataset for the mixture density network (MDN) is summarized in Algorithm \ref{alg:param_corr_time_variate_rates_LVGBM}. The target in this case is also the weighted return of the basket, but we include the basket weights as part of the MDN input.

\begin{algorithm}
\caption{Training Set Generation for local-volatility GBM with Correlated Assets}\label{alg:param_corr_time_variate_rates_LVGBM}
\begin{algorithmic}
\State \textbf{Inputs:}  Initial basket asset price $S(0)$
\State \textbf{Output:} Training dataset ${(\bfx^{m,i}, \mathbf{y}^{m,i})}_{i,m}$
\State 
\State \textbf{Step 1:} Sample $n_1$ maturity values $T^i \in [0, T]$
\State \textbf{Step 2:} For each $i = 1, \dots, n_1$:
\For{$m = 1:n_2$}
\State - Sample initial values $r_0^{m,i}$ and $q_0^{m,i}$
\State - Generate time-varying paths $r^{m,i}(t; r_0^{m,i})$ and $q^{m,i}(t; q_0^{m,i})$ for $t \in [0, T^i]$ using the CIR model
\State - Sample a lower-triangular matrix $\mathbf{L}^{m,i}$ to define the correlation structure
\State - Sample local-volatility function parameters $a_{loc}^{m,i},b_{loc}^{m,i}$ and $c_{loc}^{m,i}$
    \State - Form the parameter set:
    $$\boldsymbol{\vartheta}^{m,i}=\Big(r^{m,i}(\cdot), \big(q_j^{m,i}(\cdot)\big)_{j=1}^N, \mathbf{L}^{m,i}\Big), $$
    and
    $$\boldsymbol{\nu}^{m,i}=\Big(\Big(a_{loc,j}^{m,i},b_{loc,j}^{m,i},c_{loc,j}^{m,i}\Big)_{J=1}^N\Big)$$

    \State - Simulate $M$ independent paths of asset price, $\Big\{S^{m,i,k}(t;\boldsymbol{\vartheta}^{m,i},\boldsymbol{\nu}^{m,i})\Big\}_{k=1}^M$ over $t\in [0,T^i]$
    \State - Generate samples of the basket weights $w^{m,i}=\Big(w_1^{m,i},\cdots,w^{m,i}_N\Big)$
\EndFor
\State \textbf{Step 3:} Compute MDN targets:
$$\mathbf{y}^{m,i}=\Bigg\{y^{m,i,k} =\log\bigg( \sum_{j=1}^N w^{m,i}_j \frac{S_j^{m,i,k}\big(T^i;\;\boldsymbol{\vartheta}^{m,i},\;\boldsymbol{\nu}^{m,i}\big)}{S_j(0)}\bigg)\Bigg\}_{k=1}^M$$
\State \textbf{Step 4:} Define MDN inputs as $ \bfx^{m,i}=(\boldsymbol{\vartheta}^{m,i},\boldsymbol{\nu}^{m,i},T^i,w^{m,i})$
\end{algorithmic}
\end{algorithm}

\subsection{MDN Training Algorithm}

The training of the MDN proceeds via the following steps:\\

\textbf{Step 1. Initialization:} Set the learning rate $\eta$ and initialize the network parameters
$$\Phi = \{\cW_0,\cdots,\cW_H,\cW_{\pi},\cW_{\mu}, \cW_{\delta},\beta_0,\cdots,\beta_H,\beta_{\pi},\beta_{\mu}, \beta_{\delta}\}.$$

\textbf{Step 2. Forward Pass:} For each training pair $(\mathbf{x}^i, \mathbf{y}^i)$, compute the MDN output $$\big\{\pi_j(\mathbf{x}^i), \mu_j(\mathbf{x}^i), \delta_j(\mathbf{x}^i)\big\}_{j=1}^d.$$

\textbf{Step 3. Compute Negative $\log$-Likelihood Loss:} The network is trained by minimizing the negative $\log$-likelihood (NLL) of the conditional density $p(\mathbf{y}|\mathbf{x})$, which corresponds to the maximum likelihood estimation of the mixture model parameters.

For each data point $(\mathbf{x}^i,\mathbf{y}^i)$, the $\log$-likelihood is given by:
\begin{equation}
     \log p(\mathbf{y}^i|\mathbf{x}^i) = \sum_{k=1}^M\log \left( \sum_{j=1}^d \pi_j(\mathbf{x}^i)\times \phi_j(y^{i,k}|\mu_j(\mathbf{x}^i),\delta_j(\mathbf{x}^i))\right).
\end{equation}

To enhance numerical stability during training, especially in the case of Gaussian mixtures, we adopt the \texttt{LogSumExp} trick and express the $\log$-likelihood as:

\begin{equation}
    \log p(\mathbf{y}^i|\mathbf{x}^i) = \sum_{k=1}^M\log \left( \sum_{j=1}^d \exp{\Big( \log \pi_j(\mathbf{x}^i)+ \log\big(\phi_j(y^{i,k}|\mu_j(\mathbf{x}^i),\delta_j(\mathbf{x}^i))\big)\Big)}\right),
\end{equation}
where the $\log$-density of a univariate Gaussian component is given by:
\begin{equation}
    \log\big(\phi_j(y^{i,k}|\mu_j(\mathbf{x}^i),\delta_j(\mathbf{x}^i))\big) = -\log \;\delta_j(\mathbf{x}^i) -\frac{\log(2\pi)}{2} -\frac{1}{2} \left(\frac{y^{i,k}-\mu_j(\mathbf{x}^i)}{\delta_j(\mathbf{x}^i)}\right). 
\end{equation}

The total loss over the entire dataset is:
\begin{align}\label{NLL_MDN}
    \cL_{\text{NLL}}& = -\frac{1}{N_s} \sum_{i=1}^{N_s} \log p(\mathbf{y}^i|\mathbf{x}^i) .
\end{align}

\textbf{Step 4. Gradient Computation:} Compute the gradients $\partial \cL /\partial \Phi$ via backpropagation using the chain rule. This requires computing the partial derivatives with respect to the mixture parameters $\pi_j,\mu_j,\delta_j$.

\textbf{Step 5. Parameter Update:} Update the network parameters using an appropriate optimizer (Adam, for example).

Repeat steps $2-5$ for multiple epochs until the training loss converges. 
%%%%%%%%%%%%%%%%%%%%%%%%%%%%%%%%%%%%%%%%%%%%

\section{Numerical Experiments}\label{sec:Numerical_Experiments}

In this section, we present numerical experiments of training a mixture density network (MDN) to approximate the conditional distribution of weighted basket returns for option pricing. As mentioned earlier, numerical experiments are divided into two subsections, corresponding to two distinct volatility structures considered under Geometric Brownian motion (GBM) dynamics for the underlying assets. Each subsection includes numerical experiments that evaluate the accuracy of the learned distribution and the resulting option price with the benchmark Monte Carlo.\\

\noindent Under both volatility structures, we consider a basket of $N$ assets, each following a GBM with time-varying interest and dividend rates. The option is written on the weighted price of the basket. Each asset evolves under its own set of parameters and is correlated with others via a given correlation matrix $\mathbf{R}\in \bR^{N \times N}$, which defines the joint dynamics of the underlying $N$-dimensional Brownian motion. Our objective is to approximate the univariate conditional distribution of the weighted return of the basket at a fixed terminal time. To this end, we employ a mixture density network (MDN) with a parametric Gaussian mixture model. All parameters of the GBM model, including the basket's correlation matrix and the terminal time, constitute the parameter set of the mixture model. \\

\noindent Before detailing numerical experiments with the two volatility structures considered, we first present the metrics used to assess distributional and pricing accuracy.\\

\noindent \textbf{Measure of Approximation Accuracy:} The accuracy of the approximations produced by the MDN can be assessed in two complementary ways:

\begin{itemize}
    \item \textbf{Distributional Accuracy:} We evaluate how closely the probability distribution learned by the MDN matches the empirical distribution obtained from Monte Carlo (MC) simulations. The empirical distribution is estimated using kernel density estimation (KDE) on the MC samples. The discrepancy between the MDN-predicted density \( p_{\text{MDN}} \) and the MC-based density \( p_{\text{MC}} \) is quantified using the Kullback-Leibler (KL) divergence:
    \[
    D_{\text{KL}}(p_{\text{MC}} \,\|\, p_{\text{MDN}}) \approx \sum_i p_{\text{MC}}(x_i) \log \frac{p_{\text{MC}}(x_i)}{p_{\text{MDN}}(x_i)} \cdot \Delta x,
    \]
    where \( \{x_i\} \) is a uniform grid over the support of the distribution and \( \Delta x \) is the grid spacing.

    \item \textbf{Pricing Accuracy:} Using the MDN-predicted distribution, we can compute option prices and compare them with benchmark prices obtained via Monte Carlo simulation. The accuracy of these prices is evaluated using a Huberized relative error metric:
    \[
    \text{relative\_error} = \frac{ \left|P_{\text{MDN}} - P_{\text{MC}}\right| }{0.125\% \times P_{\text{MC}} + 0.00125},
    \]
    where \( P_{\text{MDN}} \) and \( P_{\text{MC}} \) denote the MDN-based and Monte Carlo-based option prices, respectively. The denominator is chosen to ensure stability of the metric for small values of \( P_{\text{MC}} \).
\end{itemize}

\noindent Now, we turn to the details of our first experiment, where each asset in the basket follows a Geometric Brownian motion with time-varying volatility.

\subsection{Basket of Assets Following GBM with Time-varying Volatility}

% \subsubsection{Numerical Experiments}
    We consider a basket of assets, where each asset has a predefined weight in the basket. The time-varying parameters $r(t)$, $q(t)$ and $\sigma(t)$ are simulated using the CIR model, with the corresponding parameter values and initial ranges summarized in Table \ref{tab:CIR_parameters_r_q}.

\begin{table}[H]
    \begin{center}
\begin{tabular}{ |c|c|c|c|c|c|c|c| } 
 \hline
  & $a$ & $b$ & $c$  & $x_0$\\ 
 \hline
 \hline
   $r(t)$ & 0.6  & 0.05  & 0.05 & [0.005, 0.1] \\ 
 \hline
  $q(t)$ & 0.6  & 0.03  & 0.02 & [0.005,0.1]\\
 \hline
  $\sigma(t)$ & 0.75  & 0.1  & 0.2 & [0.01,0.2]\\
  \hline
\end{tabular}
 \caption{CIR model parameters}
    \label{tab:CIR_parameters_r_q}
\end{center}
\end{table}

   To construct the training dataset, we simulate $n_1=5000$ samples of maturity $T^i;i=1,\cdots,n_1$. For each $i$, we generate further $n_2=4000$ independent samples of GBM model parameters $\boldsymbol{\vartheta}^{m,i}$ for $m=1,\cdots,n_2$. This yields a total of 20 million input samples $\bfx^{m,i}=(\boldsymbol{\vartheta}^{m,i},T^i);i=1,\cdots,n_1;m=1,\cdots,n_2$. For each input $\bfx^{m,i}$, we simulate $M=30$ independent paths of the asset price process. The corresponding target values are
   $$\mathbf{y}^{m,i}=\Bigg\{\log\bigg( \sum_{j=1}^N w_j \frac{S_j^{m,i,k}(T^i;\;\boldsymbol{\vartheta}^{m,i})}{S_j(0)}\bigg)\Bigg\}_{k=1}^M.$$ 
  Thus, each input data is associated with $M$ likelihood samples for training the MDN.

\subsubsection{Input Features} Since using the full path of time-varying parameters (e.g., $r(t), q(t)$ or $\sigma(t)$), as inputs to the MDN is impractical, we instead utilize truncated signatures of these paths. For all experiments, the signatures are truncated at level $l^{sig}=5$.

This results in a total of $2N+(1+2N)l^{sig}+2+\frac{N(N+1)}{2}$ custom input features. The input vector for the MDN takes the form:
\begin{align*}
    \bfx(\boldsymbol{\vartheta}, T)= &\Big(r(\cdot)T,\; q(\cdot)T,\;  \sigma(\cdot)T,\; \mathbf{L}\sqrt{T},\; T\Big)\\
    = &\Big(r^{mean}T,\;q_1^{mean}T,\;q_2^{mean}T,\;\sigma_1^{mean}\sqrt{T},\;\sigma_2^{mean}\sqrt{T},\; r^{sig}T,\; q_1^{sig}T,\; q_2^{sig}T,\; \sigma_1^{sig}T,\; \sigma_2^{sig}T,\;\mathbf{L}\sqrt{T}, \;T\Big),
\end{align*}
where $r^{mean}$ and $r^{sig}$ denote the mean and signature terms derived from the path of $r(t)$ over $[0,T]$, and likewise for $q(t)$ and $\sigma(t)$. Since signature values at higher levels diminish rapidly in magnitude, we normalize all signature components for each sample before using them as input features.

\subsubsection{MDN Architecture:}\label{MDN_structure_time_vol} The MDN used here is univariate and composed of 6 hidden layers, with the following number of neurons per layer: 320, 256, 256, 192, 128 and 80. The architecture uses the following activation and transformation functions:
\begin{align*}
    &\beta_0=\beta_1=\cdots=\beta_5=\beta_{\mu}=\beta_{\delta}=0,\\
     &g(\bfx)=\textit{LeakyReLU}(\bfx),\\
     &g_{\pi}(\bfx)=\textit{softmax}(\bfx),\\
     &g_{\mu}(\bfx)={\tanh}(\bfx),\\
     &g_{\delta}(\bfx)=\textit{softplus}(\bfx)*{\tanh}^2(\bfx)+\eps^0,
\end{align*}
where $\eps^0$ is a small positive constant used to ensure numerical stability in the standard deviation output. Unless 

\subsubsection{Model Training} We employ a mixture density network with $d=10$ Gaussian components. The model is trained using the $\textit{AdamW}$ optimizer. In each epoch, the full training dataset of 20 million samples is processed in mini-batches of size 100,000. After each epoch, performance is evaluated on a validation set of 5 million samples. The initial learning rate is set to $0.01$ and is adaptively reduced when validation loss stagnates over multiple epochs.

\subsubsection{Numerical Results} 
\begin{figure}[hbtp]
    \centering
    \subfloat[\label{fig:GBM_TV_corr_time_variate_1a}]{\includegraphics[width=0.3\textwidth]{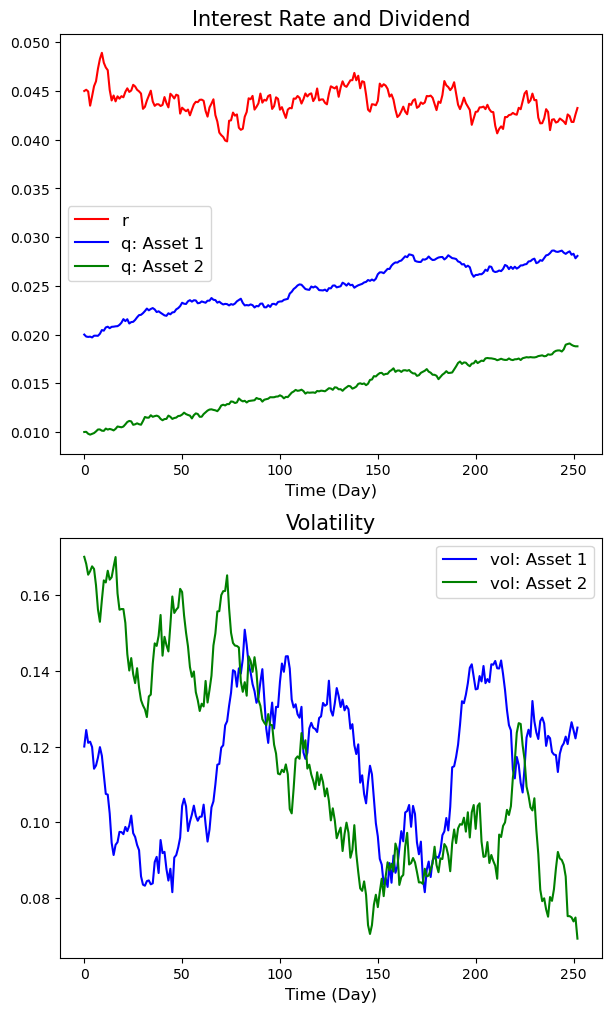}}\hfill
    \subfloat[\label{fig:GBM_TV_corr_time_variate_1b}]{\includegraphics[width=0.46\textwidth]{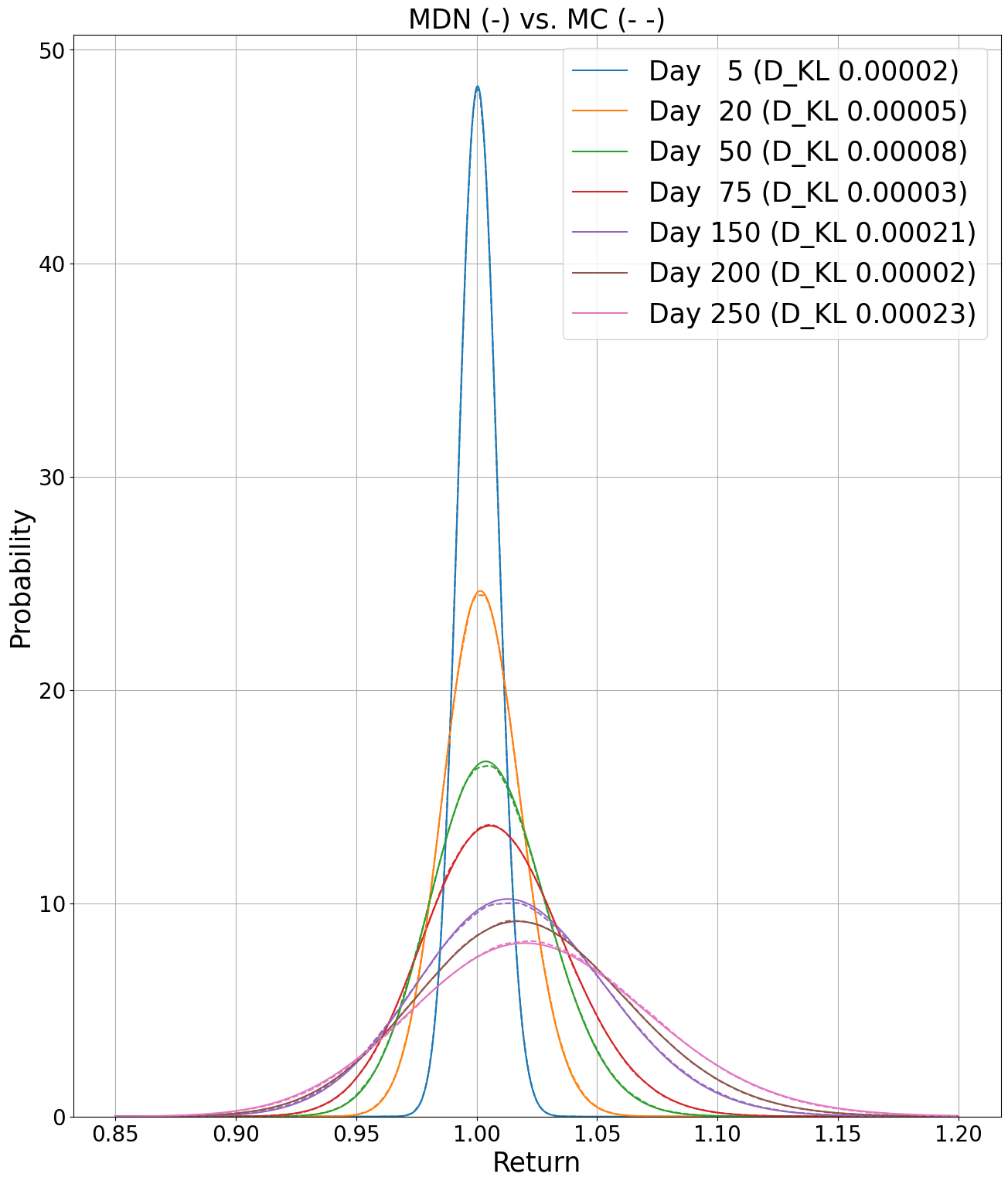}}
    \caption{(a) Time-varying interest rate, dividend yields, and volatilities. (b) MDN vs. MC distributions of average basket return with KL divergence at different maturities.}
    \label{fig:GBM_TV_N2CSign_MDN_MC_1}
\end{figure}

\textbf{2-Asset Basket:} We consider a basket of two assets ($N=2$), where each asset has equal weight $(w_1=0.5, w_2=0.5)$. For the simulation of the training dataset, we set the initial asset price as $S(0)=(S_1(0),S_2(0))=(1.0, 1.0)$. The time series of the risk-free interest rate, dividend yields, and volatilities, along with the correlation matrix, serve as inputs to the MDN. 
Assume, Figure~\ref{fig:GBM_TV_corr_time_variate_1a} presents the time series of the risk-free interest rate, along with the dividend yields and volatilities for the two assets in the basket, simulated over a one-year horizon (252 trading days). The correlation structure among these assets is specified by the correlation matrix
\[
\mathbf{R} = \begin{pmatrix}
1 & -0.7131 \\
-0.7131 & 1
\end{pmatrix}.
\]

Then, Figure~\ref{fig:GBM_TV_corr_time_variate_1b} compares the resulting estimated distributions of the average return of the basket obtained using the mixture density network (MDN) and those obtained from Monte Carlo (MC) simulation, across different maturities. The label “D\_KL” beside each maturity denotes the Kullback–Leibler (KL) divergence between the MDN and MC distributions, providing a quantitative measure of how closely the MDN replicates the actual return distribution.

\begin{figure}[hbtp]
    \centering
    \includegraphics[width=.5\textwidth]{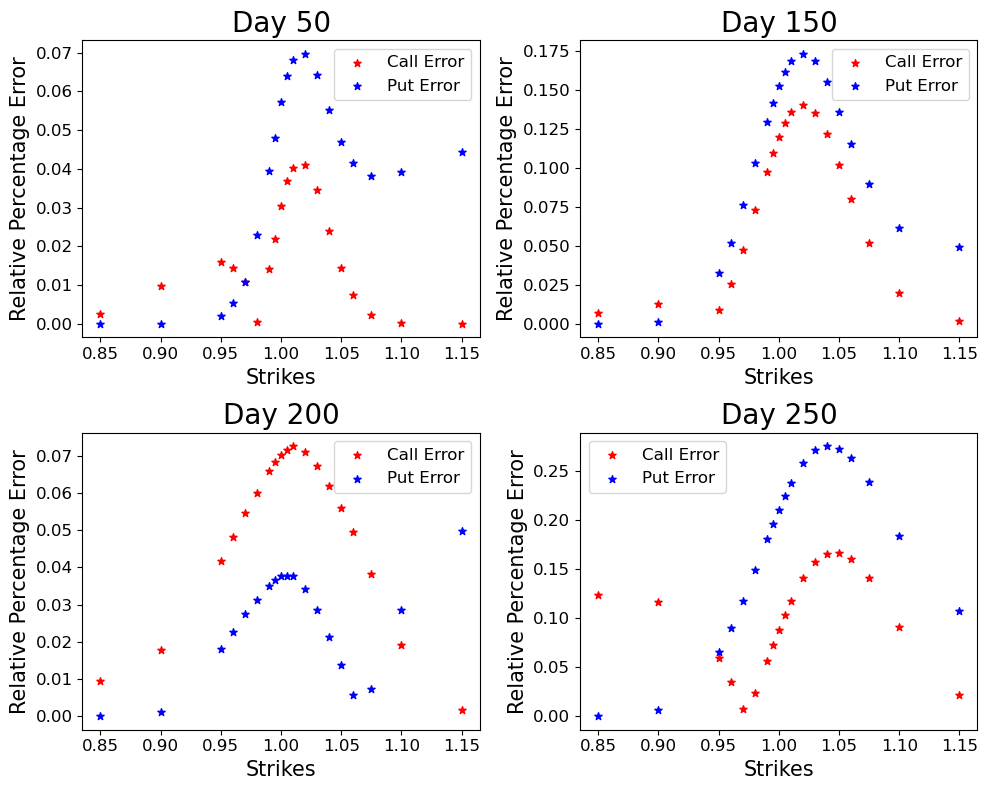}
    \caption{Relative percentage error in European call and put option prices based on MDN vs. MC pricing.}
    \label{fig:GBM_TV_N2CSign_price_error_1}
\end{figure}

Under the same market conditions, we report the relative percentage errors between option prices computed from the MDN-based distribution and those from the MC-based distribution in Figure~\ref{fig:GBM_TV_N2CSign_price_error_1}. These results are shown separately for European call and put options at various strikes and maturities, offering insights into the pricing accuracy achieved by the MDN approximation.\\

\begin{figure}[hbtp]
    \centering
    \subfloat[\label{fig:GBM_TV_corr_time_variate_2a}]{\includegraphics[width=0.3\textwidth]{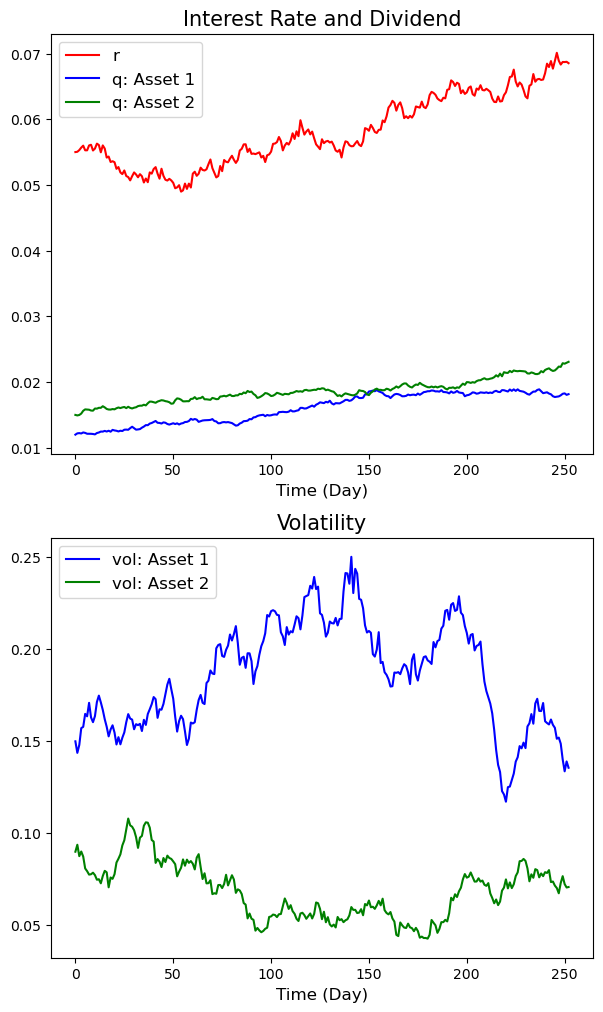}}\hfill
    \subfloat[\label{fig:GBM_TV_corr_time_variate_2b}]{\includegraphics[width=0.46\textwidth]{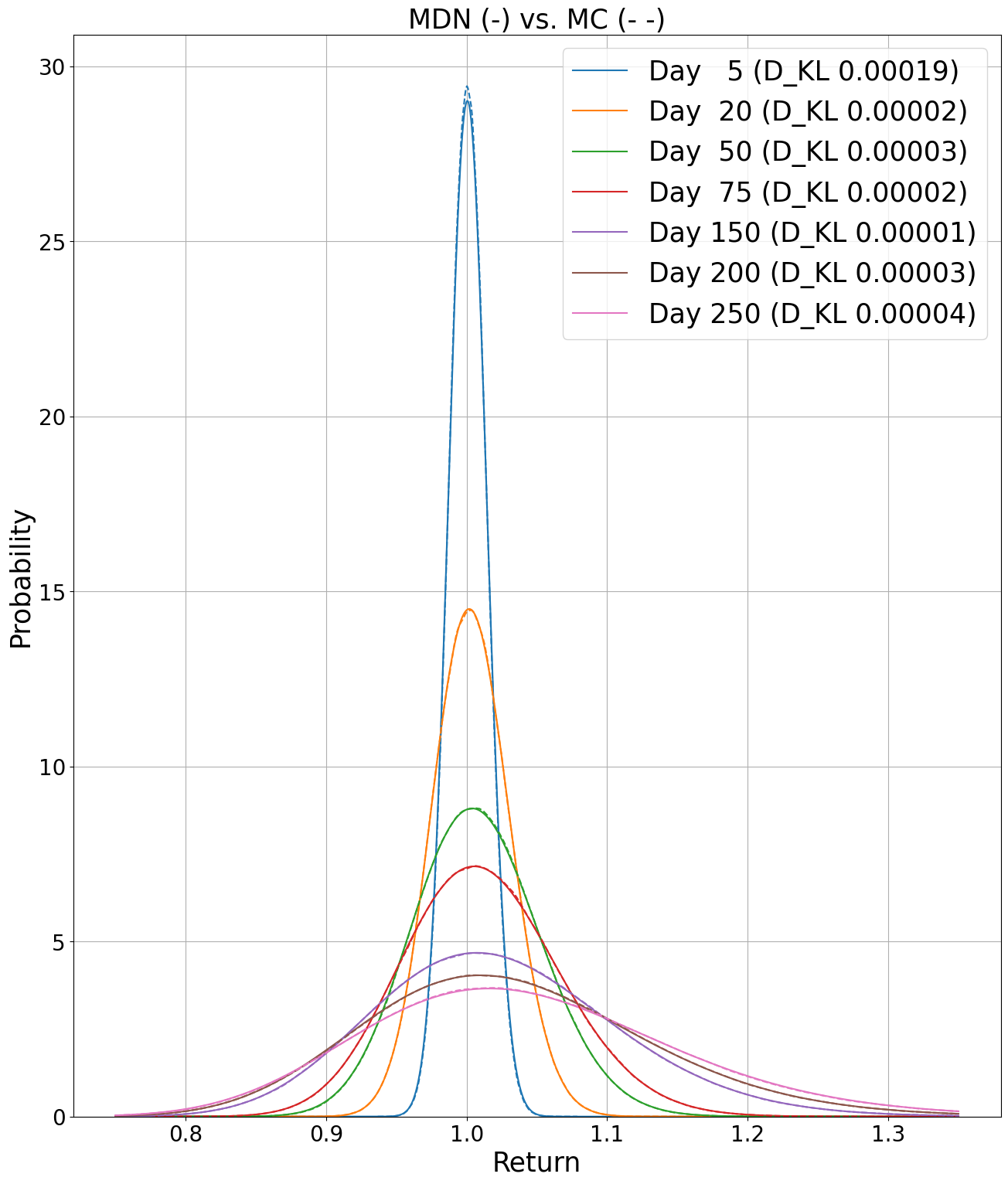}}
    \caption{(a) Time-varying interest rate, dividend yields, and volatilities. (b) MDN vs. MC distributions of average basket return with KL divergence at different maturities.}
    \label{fig:GBM_TV_N2CSign_MDN_MC_2}
\end{figure}

\begin{figure}[hbtp]
    \centering
    \includegraphics[width=.5\textwidth]{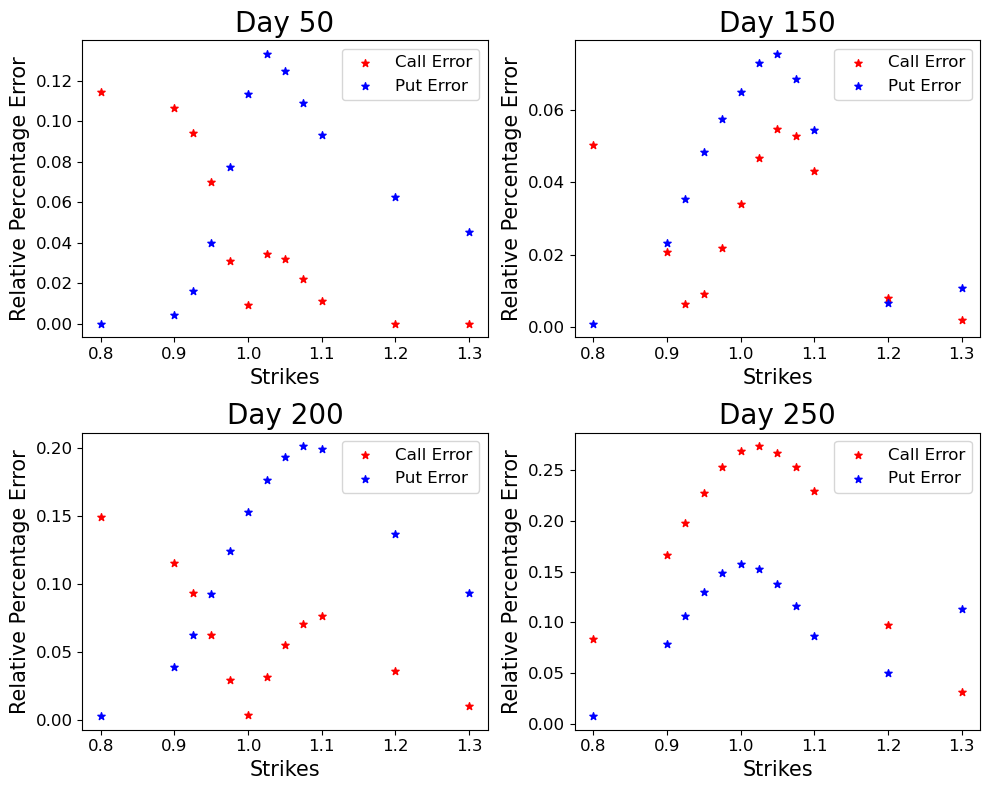}
    \caption{Relative percentage error in European call and put option prices based on MDN vs. MC pricing.}
    \label{fig:GBM_TV_N2CSign_price_error_2}
\end{figure}

\noindent In a second scenario, the correlation matrix is changed to
\[
\mathbf{R} = \begin{pmatrix}
1 & 0.219 \\
0.219 & 1
\end{pmatrix},
\]
while the time series of the risk-free rate, dividend yields, and volatilities are given in Figure~\ref{fig:GBM_TV_corr_time_variate_2a}. Figure~\ref{fig:GBM_TV_corr_time_variate_2b} compares the corresponding MDN and MC return distributions, with the associated KL divergences. The resulting relative percentage pricing errors for European options are reported in Figure~\ref{fig:GBM_TV_N2CSign_price_error_2}.

\begin{figure}[hbtp]
    \centering
    \subfloat[\label{fig:GBM_TV_corr_time_variate_1a_N3}]{\includegraphics[width=0.3\textwidth]{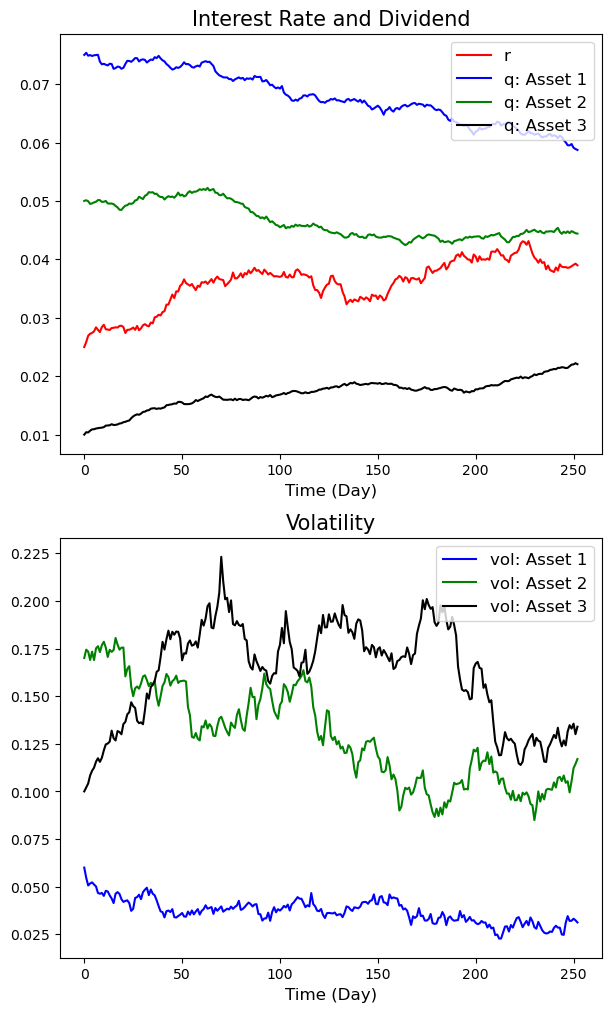}}\hfill
    \subfloat[\label{fig:GBM_TV_corr_time_variate_1b_N3}]{\includegraphics[width=0.46\textwidth]{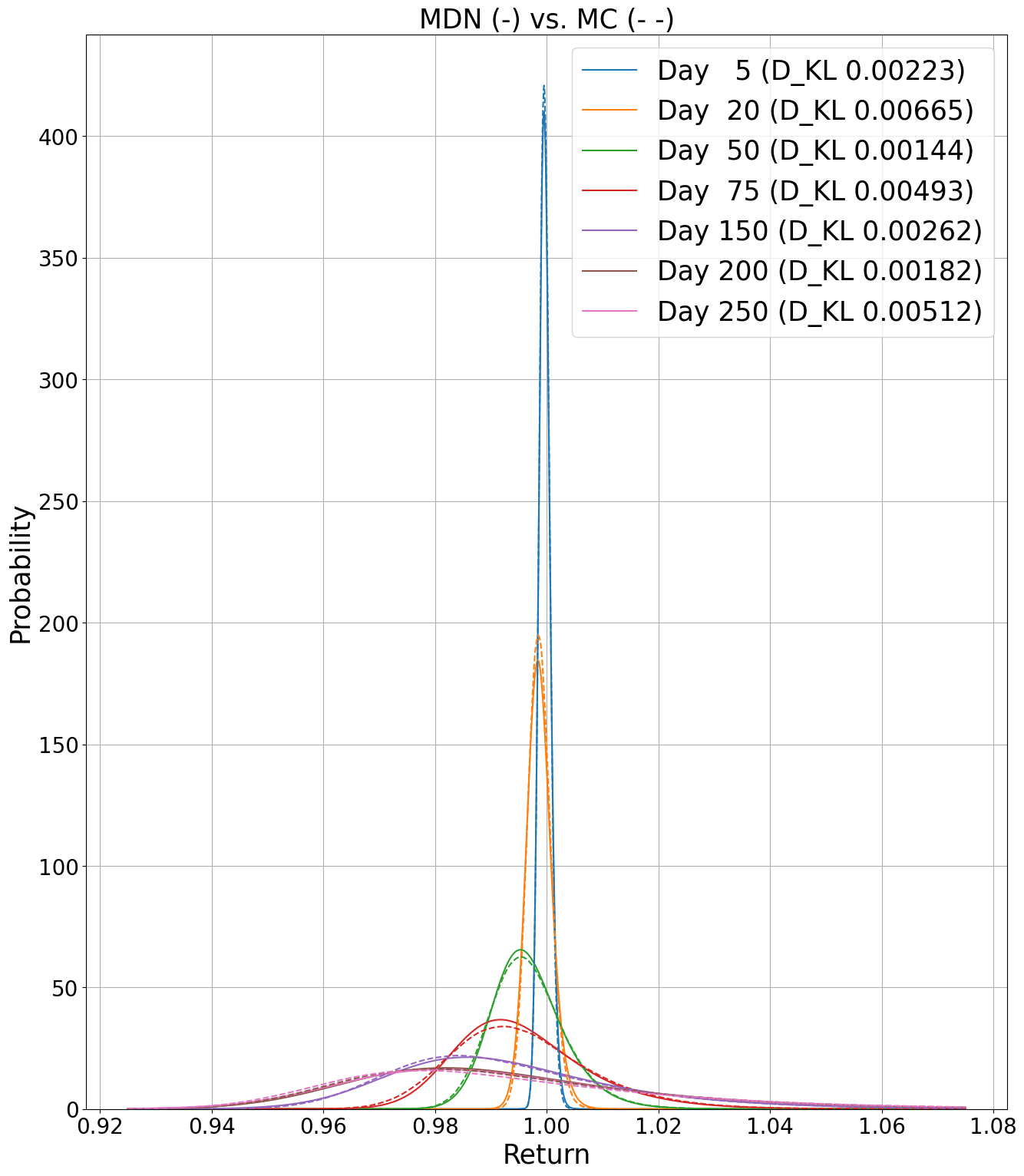}}
    \caption{(a) Time-varying interest rate, dividend yields, and volatilities. (b) MDN vs. MC distributions of average basket return with KL divergence at different maturities.}
    \label{fig:GBM_TV_N3CSign_MDN_MC_1}
\end{figure}

\FloatBarrier
\noindent \textbf{3-Asset Basket:} We consider a basket of three assets ($N=3$), where assets has weight $(w_1=0.33, w_2=0.33, w_3=0.34)$. For the simulation of the training dataset, we set the initial asset price as $S(0)=(S_1(0),S_2(0),S_3(0))=(1.0, 1.0, 1.0)$. We have used almost the same MDN structure as mentioned in Section \ref{MDN_structure_time_vol}, but with $g_{\mu}(\bfx)=\bfx+\tanh{\bfx}$. Figure~\ref{fig:GBM_TV_corr_time_variate_1a_N3} presents the time series of the risk-free interest rate, along with the dividend yields and volatilities for the three assets in the basket. The correlation structure among these assets is specified by the correlation matrix
\[
\mathbf{R} = \begin{pmatrix}
1 & -0.9968 & 0.9770 \\
-0.9968 & 1 & -0.9862 \\
0.9770 & -0.9862 & 1
\end{pmatrix}.
\]

\begin{figure}[hbtp]
    \centering
    \includegraphics[width=.5\textwidth]{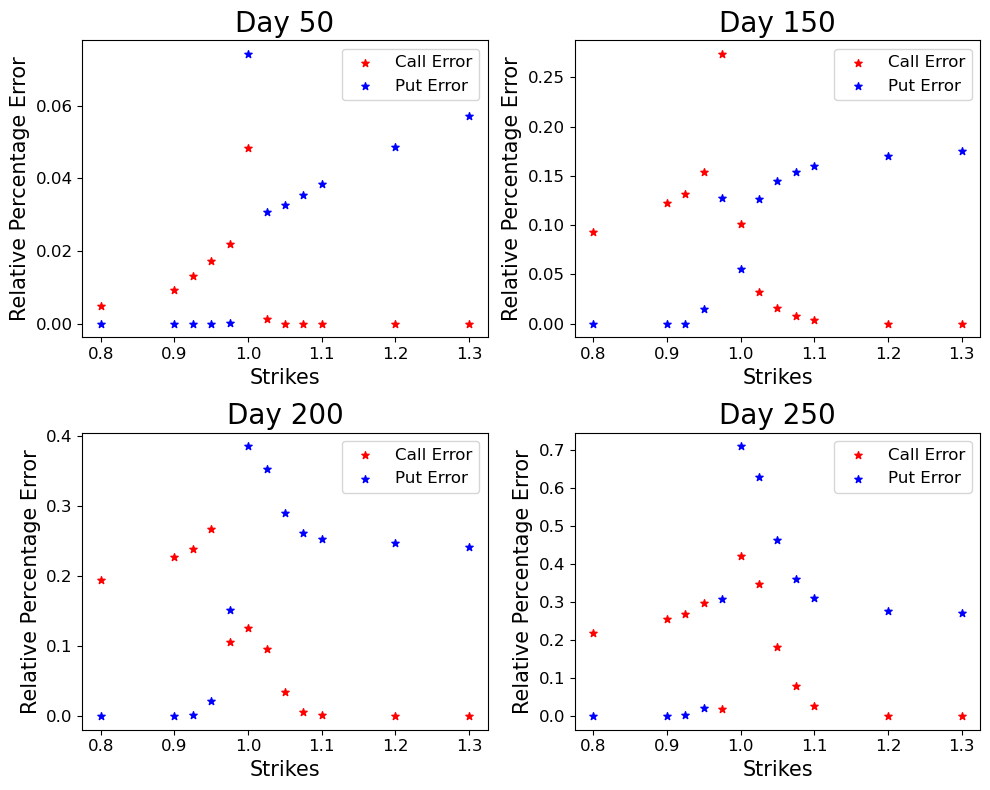}
    \caption{Relative percentage error in European call and put option prices based on MDN vs. MC pricing.}
    \label{fig:GBM_TV_N3CSign_price_error_1}
\end{figure}
Then, Figure~\ref{fig:GBM_TV_corr_time_variate_1b_N3} compares the corresponding MDN and MC return distributions, with the associated KL divergences. The resulting relative percentage pricing errors for European options are reported in Figure~\ref{fig:GBM_TV_N3CSign_price_error_1}. These results are shown separately for European call and put options at various strikes and maturities, offering insights into the pricing accuracy achieved by the MDN approximation.\\
\FloatBarrier

\begin{figure}[hbtp]
    \centering
    \subfloat[\label{fig:GBM_TV_corr_time_variate_2a_N3}]{\includegraphics[width=0.3\textwidth]{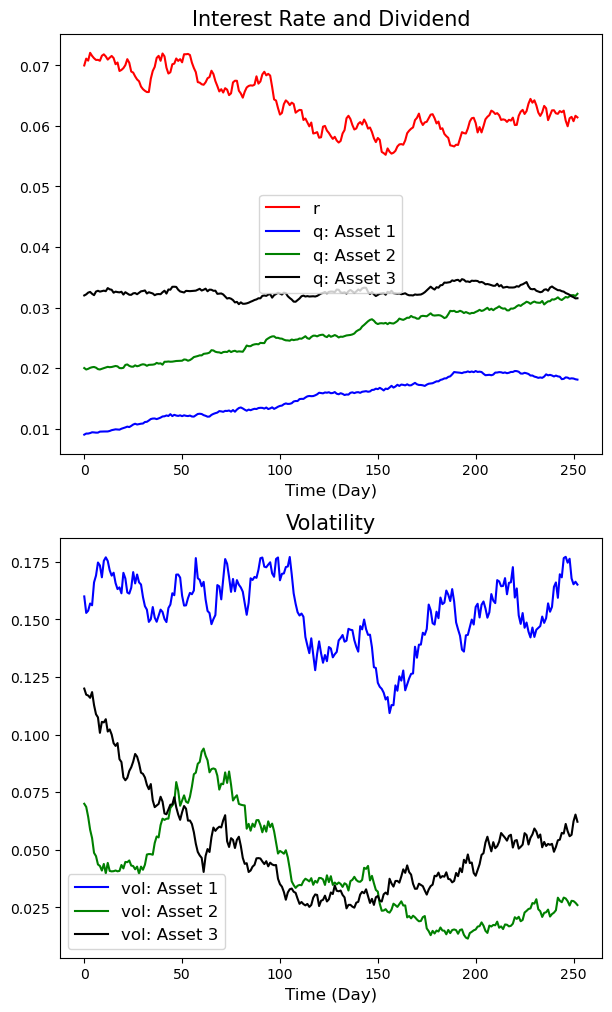}}\hfill
    \subfloat[\label{fig:GBM_TV_corr_time_variate_2b_N3}]{\includegraphics[width=0.46\textwidth]{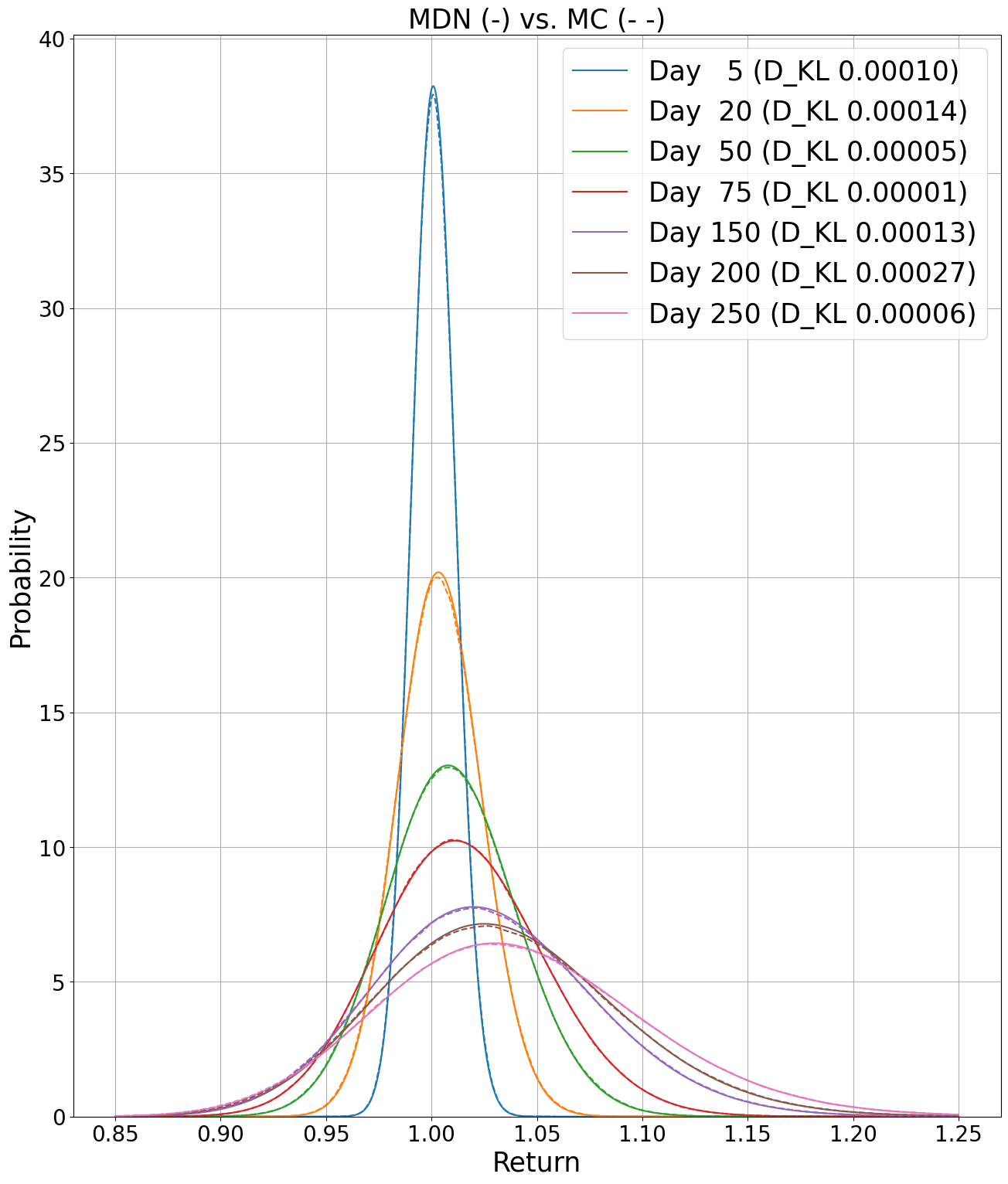}}
    \caption{(a) Time-varying interest rate, dividend yields, and volatilities. (b) MDN vs. MC distributions of average basket return with KL divergence at different maturities.}
    \label{fig:GBM_TV_N3CSign_MDN_MC_2}
\end{figure}

\begin{figure}[hbtp]
    \centering
    \includegraphics[width=.5\textwidth]{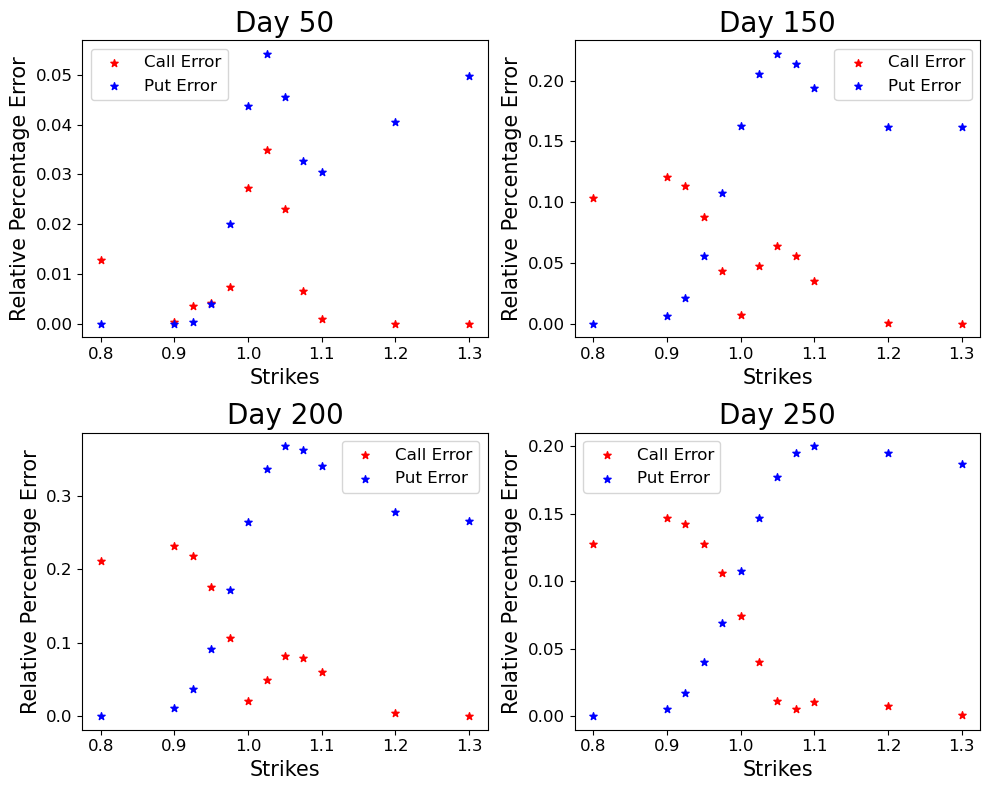}
    \caption{Relative percentage error in European call and put option prices based on MDN vs. MC pricing.}
    \label{fig:GBM_TV_N3CSign_price_error_2}
\end{figure}

\noindent In a second scenario, the correlation matrix is changed to
\[
\mathbf{R} = \begin{pmatrix}
1 & 0.5138 & -0.2963\\
0.5138 & 1 &  0.6030\\
-0.2963 & 0.6030 & 1
\end{pmatrix},
\]
while the time series of the risk-free rate, dividend yields, and volatilities are given in Figure~\ref{fig:GBM_TV_corr_time_variate_2a_N3}. Figure~\ref{fig:GBM_TV_corr_time_variate_2b_N3} compares the corresponding MDN and MC return distributions, with the associated KL divergences. The resulting relative percentage pricing errors for European options are reported in Figure~\ref{fig:GBM_TV_N3CSign_price_error_2}.

\FloatBarrier

\begin{figure}[hbtp]
    \centering
    \subfloat[\label{fig:GBM_TV_corr_time_variate_1a_N4}]{\includegraphics[width=0.3\textwidth]{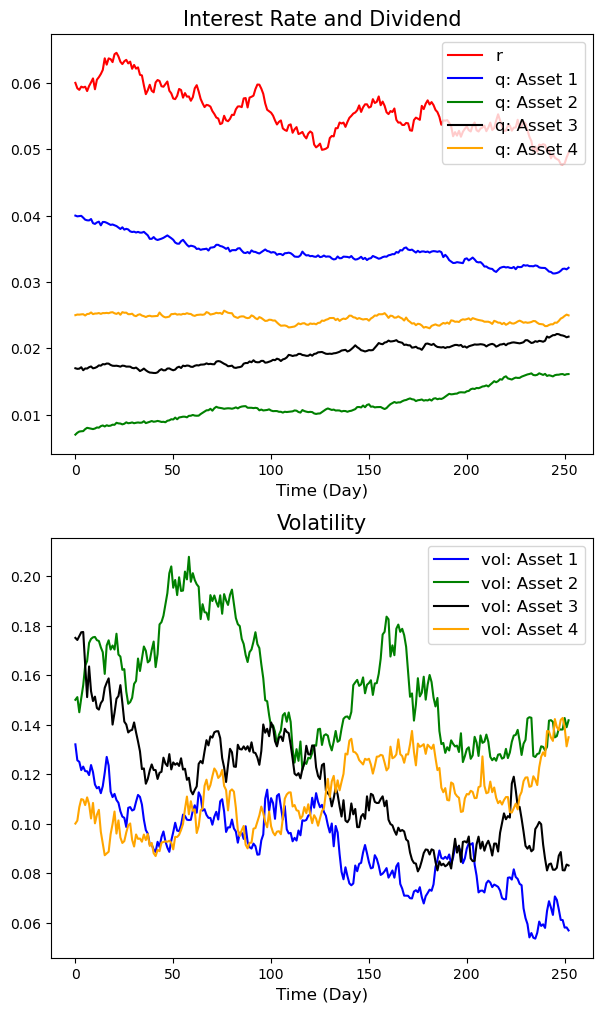}}\hfill
    \subfloat[\label{fig:GBM_TV_corr_time_variate_1b_N4}]{\includegraphics[width=0.46\textwidth]{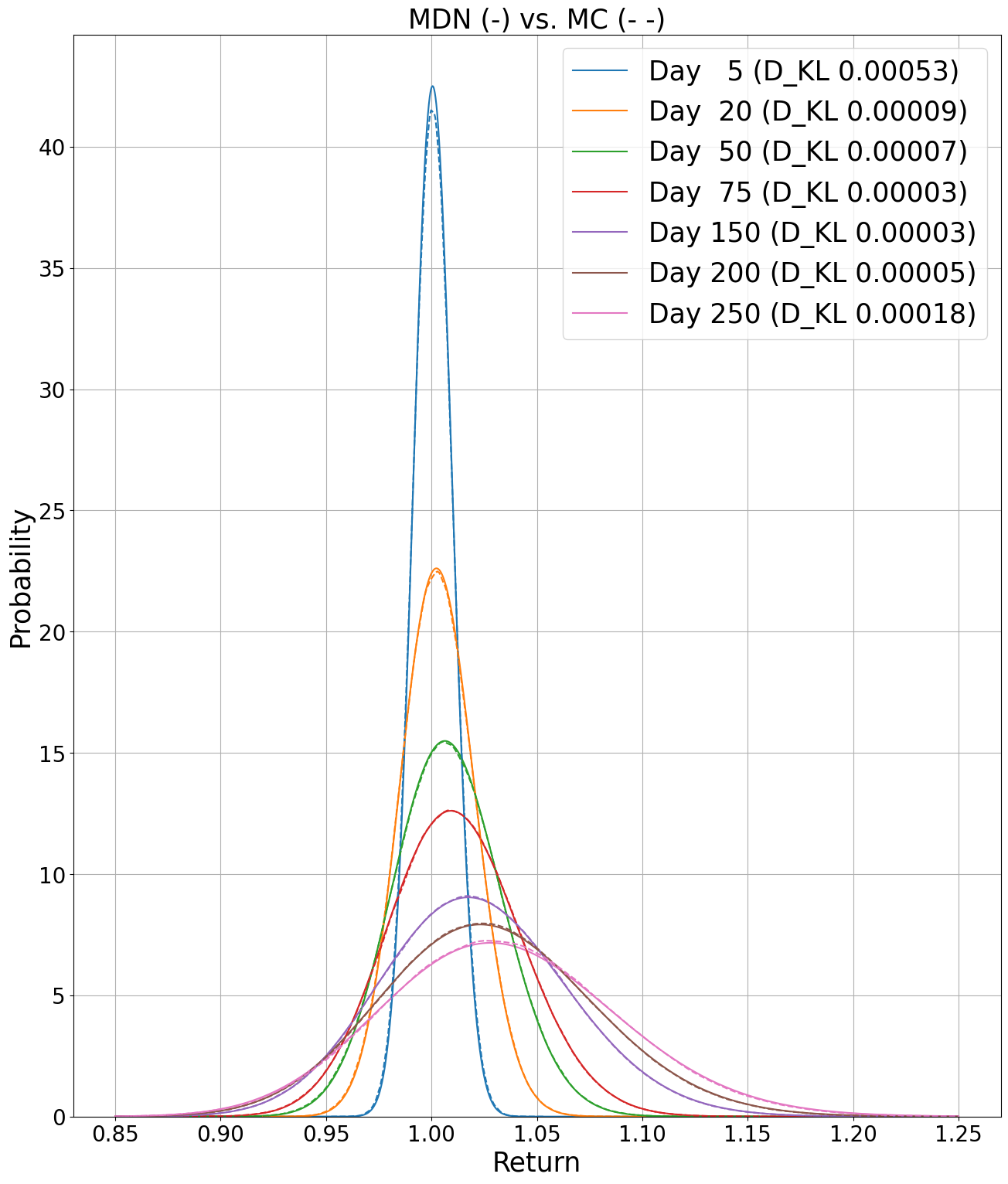}}
    \caption{(a) Time-varying interest rate, dividend yields, and volatilities. (b) MDN vs. MC distributions of average basket return with KL divergence at different maturities.}
    \label{fig:GBM_TV_N4CSign_MDN_MC_1}
\end{figure}

\noindent \textbf{4-Asset Basket:} We consider a basket of four assets ($N=4$), where each assets has same weight $(w_1=0.25, w_2=0.25, w_3=0.25, w_4 =0.25)$. For the simulation of the training dataset, we set the initial asset price as $S(0)=(S_1(0),S_2(0),S_3(0),S_4(0))=(1.0, 1.0, 1.0, 1.0)$. We have used the same MDN structure as the 3-Asset Basket case. Figure~\ref{fig:GBM_TV_corr_time_variate_1a_N4} presents the time series of the risk-free interest rate, along with the dividend yields and volatilities for the three assets in the basket. The correlation structure among these assets is specified by the correlation matrix
\[
\mathbf{R} = \begin{pmatrix}
1 & -0.6663 & 1 & 0.1460 \\
-0.6663 & 1 & -0.6652 & 0.0308 \\
1 & -0.6652 & 1 &  0.1386\\
 0.1460 & 0.0308 &  0.1386  &1 
\end{pmatrix}.
\]

\begin{figure}[hbtp]
    \centering
    \includegraphics[width=.5\textwidth]{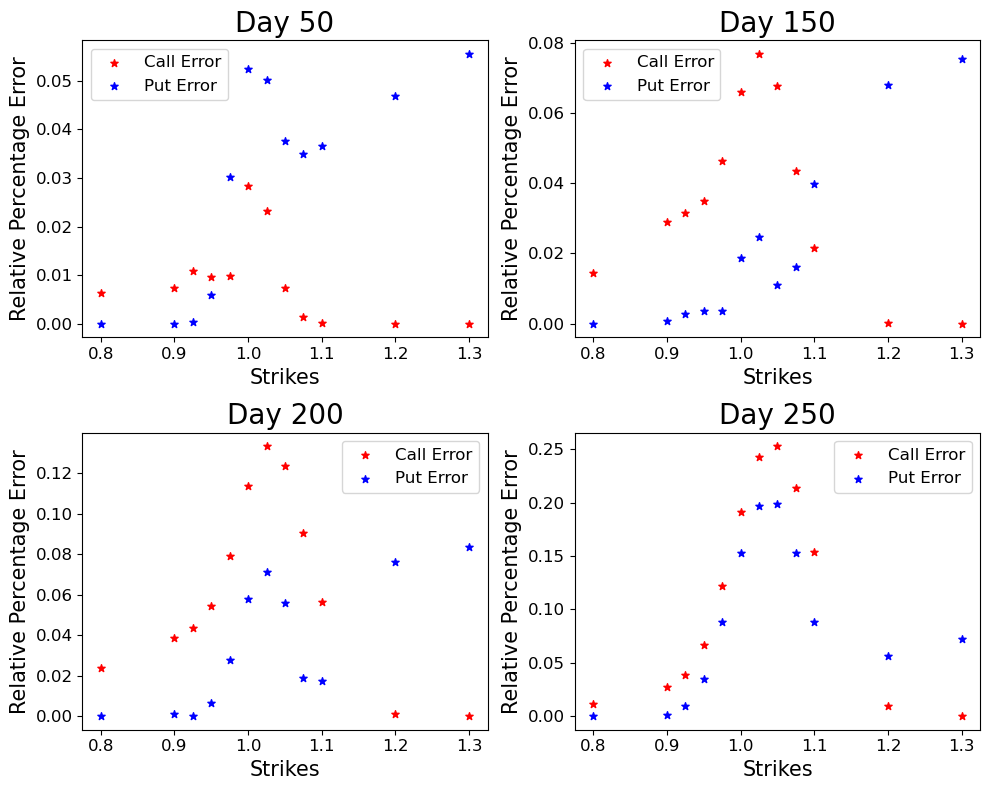}
    \caption{Relative percentage error in European call and put option prices based on MDN vs. MC pricing.}
    \label{fig:GBM_TV_N4CSign_price_error_1}
\end{figure}

\begin{figure}[hbtp]
    \centering
    \subfloat[\label{fig:GBM_TV_corr_time_variate_2a_N4}]{\includegraphics[width=0.3\textwidth]{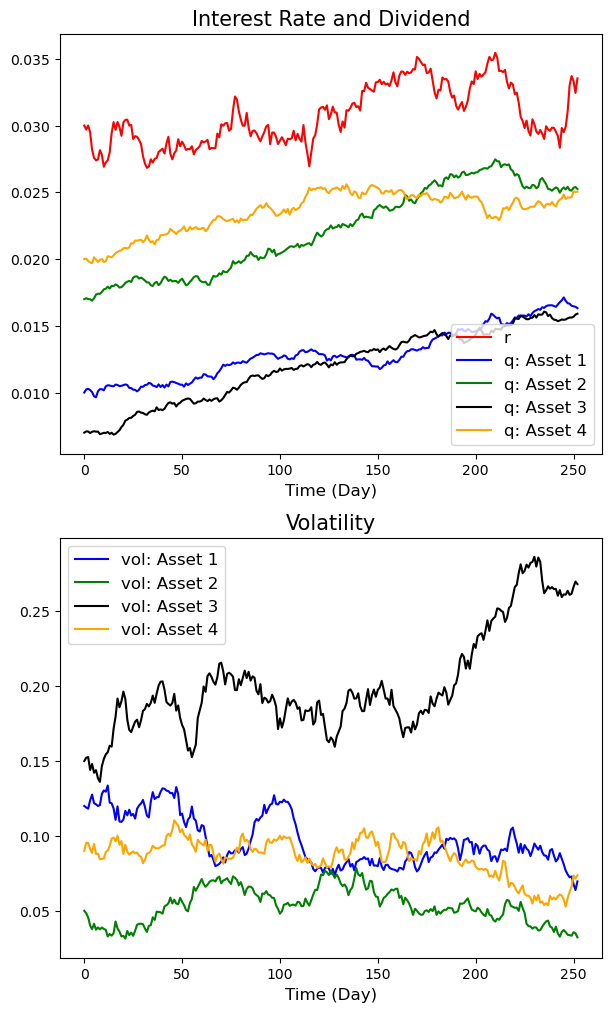}}\hfill
    \subfloat[\label{fig:GBM_TV_corr_time_variate_2b_N4}]{\includegraphics[width=0.46\textwidth]{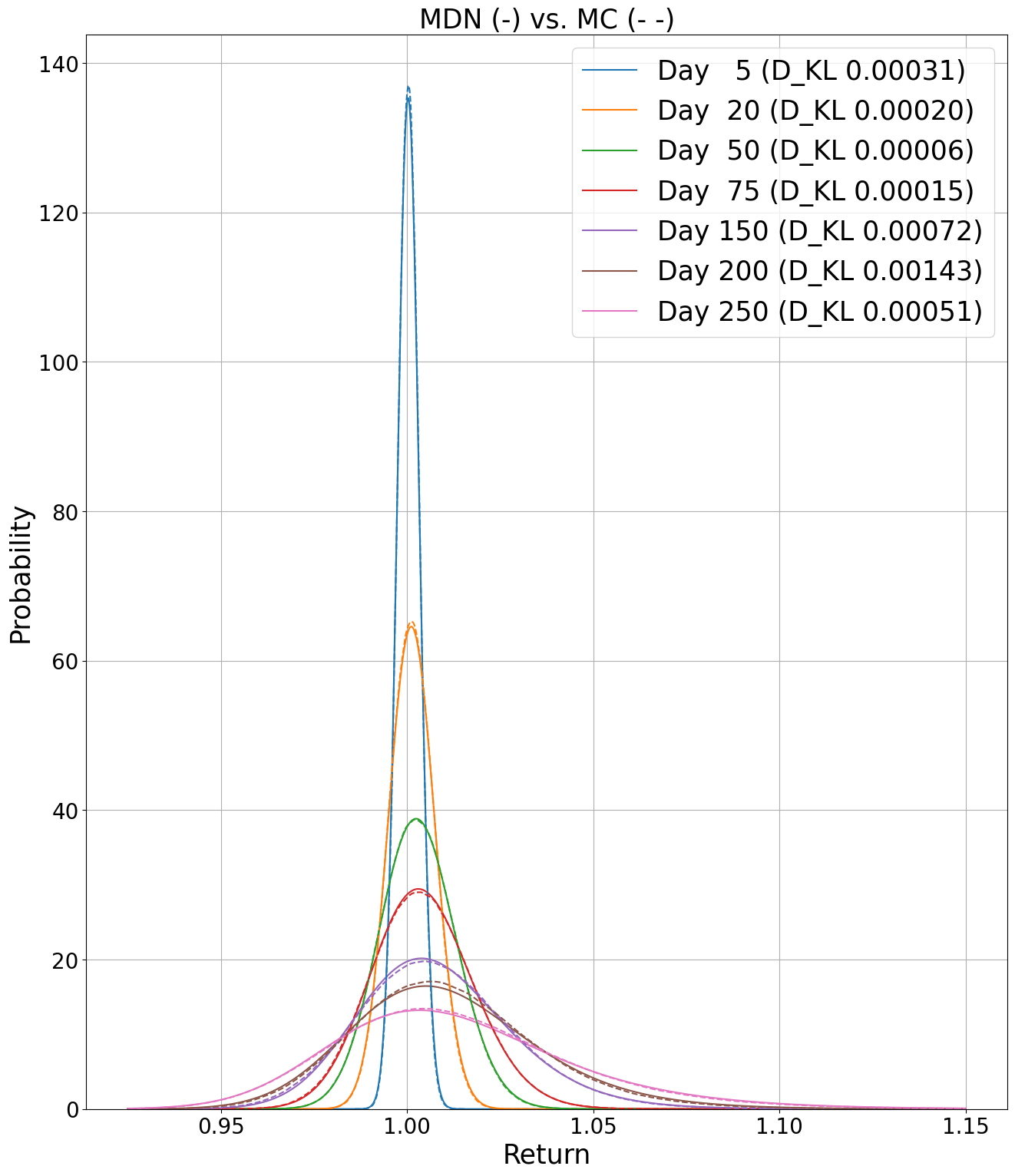}}
    \caption{(a) Time-varying interest rate, dividend yields, and volatilities. (b) MDN vs. MC distributions of average basket return with KL divergence at different maturities.}
    \label{fig:GBM_TV_N4CSign_MDN_MC_2}
\end{figure}

\begin{figure}[hbtp]
    \centering
    \includegraphics[width=.5\textwidth]{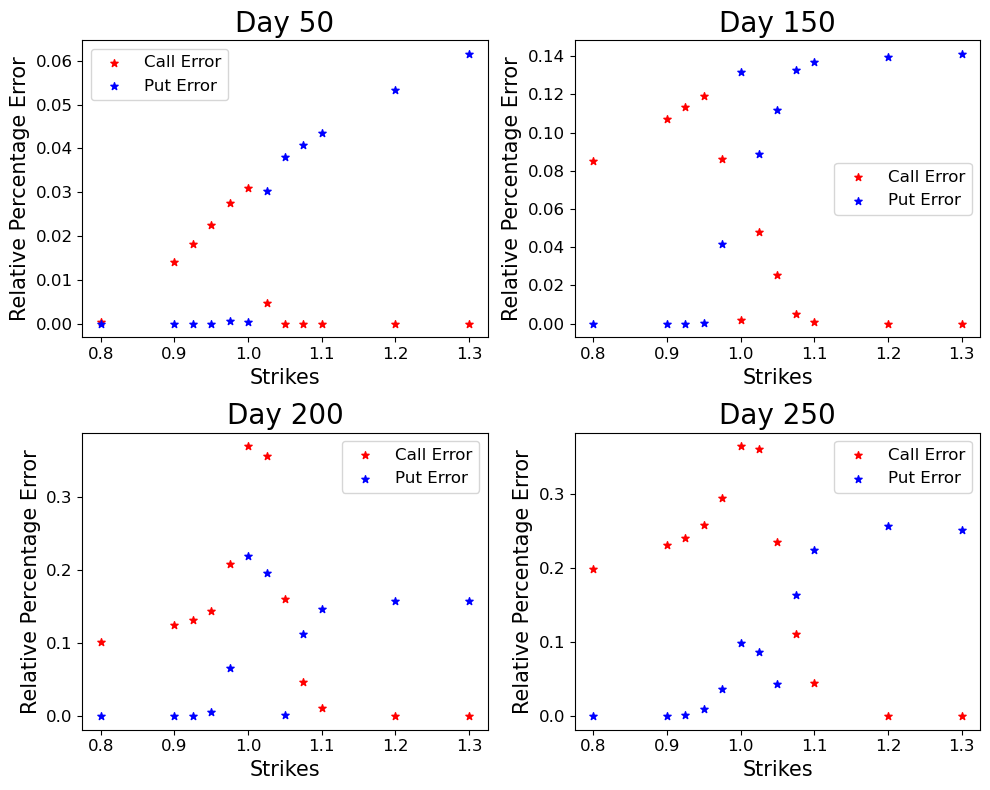}
    \caption{Relative percentage error in European call and put option prices based on MDN vs. MC pricing.}
    \label{fig:GBM_TV_N4CSign_price_error_2}
\end{figure}

Then, Figure~\ref{fig:GBM_TV_corr_time_variate_1b_N4} compares the corresponding MDN and MC return distributions, with the associated KL divergences. The resulting relative percentage pricing errors for European options are reported in Figure~\ref{fig:GBM_TV_N4CSign_price_error_1}. These results are shown separately for European call and put options at various strikes and maturities, offering insights into the pricing accuracy achieved by the MDN approximation.\\
\FloatBarrier
\noindent In a second scenario, the correlation matrix is changed to
\[
\mathbf{R} = \begin{pmatrix}
1 & -0.9780 & -0.8317 & 0.8888 \\
-0.9780 & 1 & 0.9164 & -0.9471 \\
-0.8317 & 0.9164 & 1 &  -0.8868\\
 0.8888 & -0.9471 &  -0.8868  &1 
\end{pmatrix}.
\]
while the time series of the risk-free rate, dividend yields, and volatilities are given in Figure~\ref{fig:GBM_TV_corr_time_variate_2a_N4}. Figure~\ref{fig:GBM_TV_corr_time_variate_2b_N4} compares the corresponding MDN and MC return distributions, with the associated KL divergences. The resulting relative percentage pricing errors for European options are reported in Figure~\ref{fig:GBM_TV_N4CSign_price_error_2}.

\FloatBarrier
These numerical results demonstrate that the MDN can effectively learn the conditional distribution over the parameter space with high accuracy, enabling accurate option pricing across a range of market conditions. This highlights the practical value of the approach: once trained, the MDN can efficiently approximate option prices across varying scenarios. Moreover, the use of truncated path signatures proves effective in capturing the dependence of the target distribution on time-varying model parameters.
\\

\noindent Now, we lay out the details of our second experiment, in which each asset in the basket follows a Geometric Brownian Motion with local volatility.

\subsection{Basket of Assets Following GBM with Local-Volatility}

    We consider a basket consisting of two assets ($N=2$) with the initial prices set to $S(0)=(S_1(0),S_2(0))=(1.0, 1.0)$ for the simulation of the training dataset. The time-varying parameters $r(t)$ and $q(t)$ are simulated using the CIR model as described previously, with parameter values and initial ranges summarized in Table \ref{tab:CIR_parameters_r_q}.
    
    To construct the training dataset, we simulate $n_1=5000$ samples of maturity $T^i;i=1,\cdots,n_1$. For each maturity $T^i$, we generate further $n_2=4000$ independent samples of:
\begin{itemize}

    \item GBM model parameters $\boldsymbol{\vartheta}^{m,i}$, 
    \item local-volatility function parameters $\boldsymbol{\nu}^{m,i}$ and 
    \item basket weights $w^{m,i}$,
    \end{itemize}
    for $m=1,\cdots,n_2$. This yields a total of 20 million input samples of the form 
    $$\bfx^{m,i}=(\boldsymbol{\vartheta}^{m,i},\boldsymbol{\nu}^{m,i},T^i,w^{m,i});i=1,\cdots,n_1;m=1,\cdots,n_2.$$ 
    For each input $\bfx^{m,i}$, we simulate $M=30$ independent paths of the asset price process, generating the associated target outputs:

    $$\mathbf{y}^{m,i}=\Bigg\{y^{m,i,k} =\log\bigg( \sum_{j=1}^N w^{m,i}_j \frac{S_j^{m,i,k}\big(T^i;\;\boldsymbol{\vartheta}^{m,i},\;\boldsymbol{\nu}^{m,i}\big)}{S_j(0)}\bigg)\Bigg\}_{k=1}^M.$$
Thus, each input vector $\bfx^{m,i}$ is paired with $M$ likelihood samples for training the mixture density network.

\subsubsection{Local-volatility Function $\sigma_L(\cdot)$} We adopt the following functional form for the local volatility: 
\begin{equation}
    \sigma_L(x)=c_{loc}\Big((x-a_{loc})^2 +c_{loc}\Big)^{b_{loc}},
\end{equation}
where the parameters $a_{loc}, b_{loc}, c_{loc}$ vary within the ranges:
\begin{align*}
    &a_{loc}\in[0.5,1.5],\\
    &b_{loc}\in[0.05,0.5],\\
    &c_{loc}\in[0.05,0.4].
\end{align*}
We denote the set of local volatility parameters for all assets as $\boldsymbol{\nu} = \big(a_{loc,j},b_{loc,j},c_{loc,j}\big)_{j=1}^N$. It is important to note that this choice of functional form is not driven by modelling assumptions or calibration to market data. Instead, we aim to assess whether a mixture density network (MDN) can learn the distributional features of the average basket return from local volatility parameters.
    
\subsubsection{Input Features} The input to the mixture density network (MDN) consists of custom-engineered features derived from the model parameters, including truncated path signatures of the time-varying rates. Specifically, we truncate signatures at level $l_{sig}=5.$ This will result in a total of $5N+(1+N)l^{sig}+2+\frac{N(N+1)}{2}$ input features. The final input vector for the MDN takes the form:
\begin{align*}
    \bfx(\boldsymbol{\vartheta},\boldsymbol{\nu}, T,w)= &\Big(wT,r(\cdot)T,\; q(\cdot)T,\; a_{loc}T,\; b_{loc} T, \;c_{loc} T,\; \mathbf{L}\sqrt{T},\; T\Big)\\
    = &\Big(w_1T, w_2T, r^{mean}T,\;q_1^{mean}T,\;q_2^{mean}T,\; r^{sig}T,\; q_1^{sig}T,\; q_2^{sig}T,\; a_{loc,1}T,\;a_{loc,2}^T,\; b_{loc,1} T, \;b_{loc,2} T, \\
    &\;c_{loc,1} T,\;c_{loc,2} T,\;  \mathbf{L}\sqrt{T}, \;T\Big),
\end{align*}
where $r^{mean}$ and $r^{sig}$ denote the mean and signature terms derived from the path of $r(t)$ over $ t\in[0,T]$, and likewise for $q(t)$.

\subsubsection{MDN Architecture} The univariate MDN employed in this setting consists of 6 hidden layers, with the following number of neurons per layer: 320, 256, 256, 192, 128 and 80, respectively. The activation and transformation functions used in the architecture are specified as follows:
\begin{align*}
    &\beta_0=\beta_1=\cdots=\beta_5=\beta_{\mu}=\beta_{\delta}=0,\\
     &g(\bfx)=\textit{LeakyReLU}(\bfx),\\
     &g_{\pi}(\bfx)=\textit{softmax}(\bfx),\\
     &g_{\mu}(\bfx)=\bfx,\\
     &g_{\delta}(\bfx)=\textit{softplus}(\bfx)*{\tanh}^2(\bfx)+\eps^0,
\end{align*}
where $\eps^0$ is a small positive constant used to ensure numerical stability in the standard deviation output.

This configuration is almost the same as in our previous case, the only exception is that the activation function $g_{\mu}$ doesn't use the $\tanh$ function, instead we use the identity function. This reflects the flexibility in architectural design—if carefully designed, other architectural choices may also work well.

\subsubsection{Model Training} The training procedure of the MDN follows the same method as in the previous case. Specifically, we employ a mixture density network with $d=10$ Gaussian components and train the model using the $\textit{AdamW}$ optimizer. In each epoch, the whole training dataset of 20 million samples is processed in mini-batches of size 100,000. After each epoch, performance is evaluated on a validation set of 5 million samples.

The initial learning rate is set to $0.01$ and is adaptively reduced when validation loss stagnates over multiple epochs. While the overall training strategy mirrors the previous case, it is worth emphasizing that tuning the learning rate upon stagnation in the validation loss involves some empirical adjustment. In practice, identifying when and how much to adjust the learning rate remains partly heuristic and can significantly impact training efficiency and convergence.

\subsubsection{Numerical Results} The inputs to the MDN consist of the correlation matrix of the asset basket, the local-volatility function parameters for each asset, time series of the risk-free interest rate and dividend yields, and the basket weights. We assume the correlation matrix
\[
\mathbf{R} = \begin{pmatrix}
1 & 0.4868 \\
0.4868 & 1
\end{pmatrix},
\]
and the local-volatility function parameters for the two assets as follows
$$a_{loc}=[1.155,0.95],\;b_{loc}=[0.263,0.387],\;c_{loc}=[0.077,0.145].$$ 

\begin{figure}[hbtp]
    \centering
    % Top row: Single figure
    \subfloat[\label{fig:GBM_LV_corr_time_variate_1a}]{%
        \includegraphics[width=0.3\textwidth]{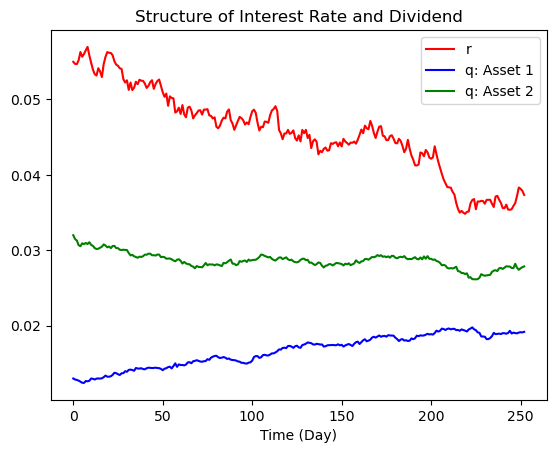}
    }

    \vspace{1ex} % vertical space between rows

    % Bottom row: Three side-by-side
    \subfloat[$w_1=0.25, w_2=0.75$\label{fig:GBM_LV_corr_time_variate_1b}]{%
        \includegraphics[width=0.3\textwidth]{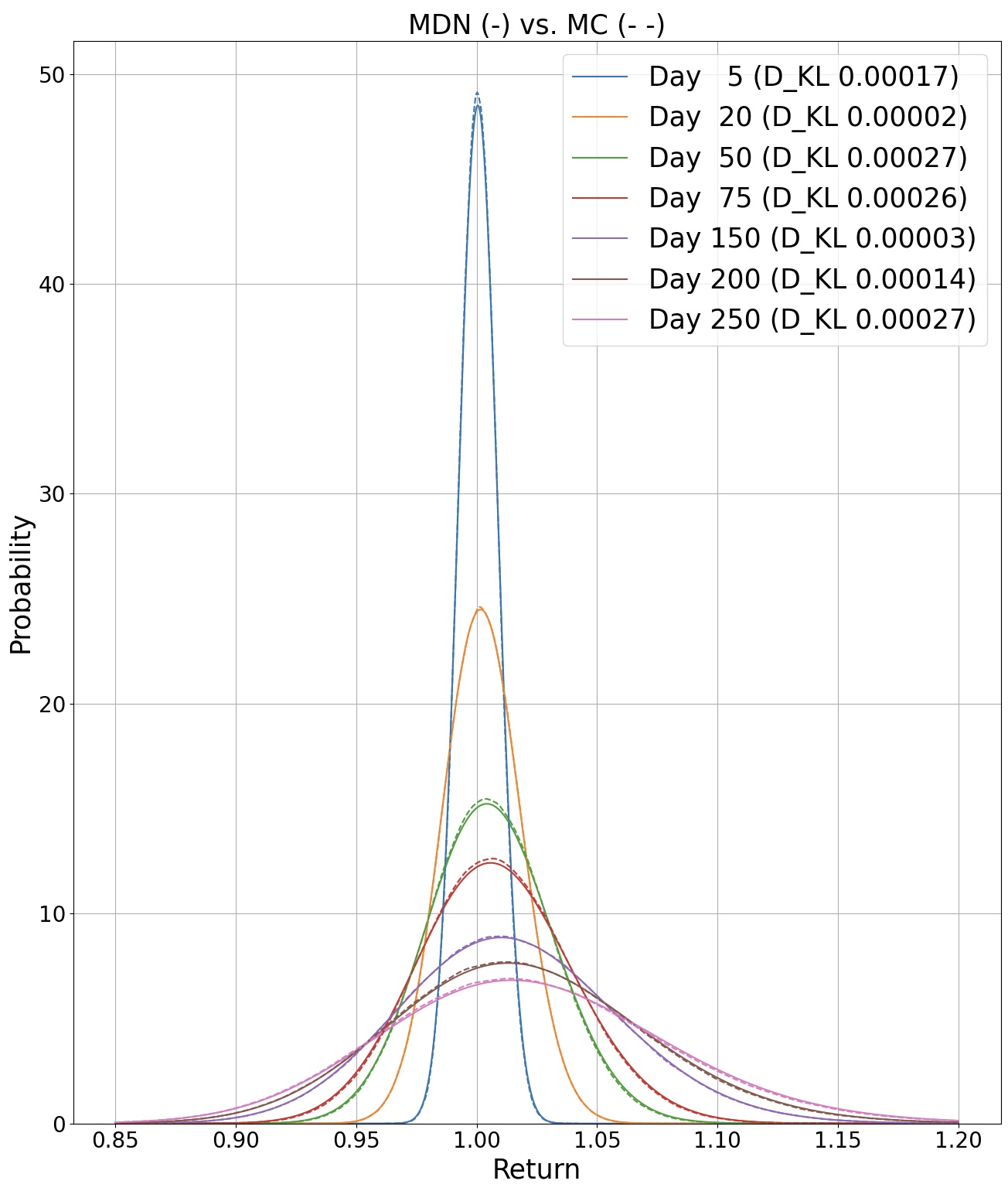}
    }\hfill
    \subfloat[$w_1=0.5, w_2=0.5$\label{fig:GBM_LV_corr_time_variate_1c}]{%
        \includegraphics[width=0.3\textwidth]{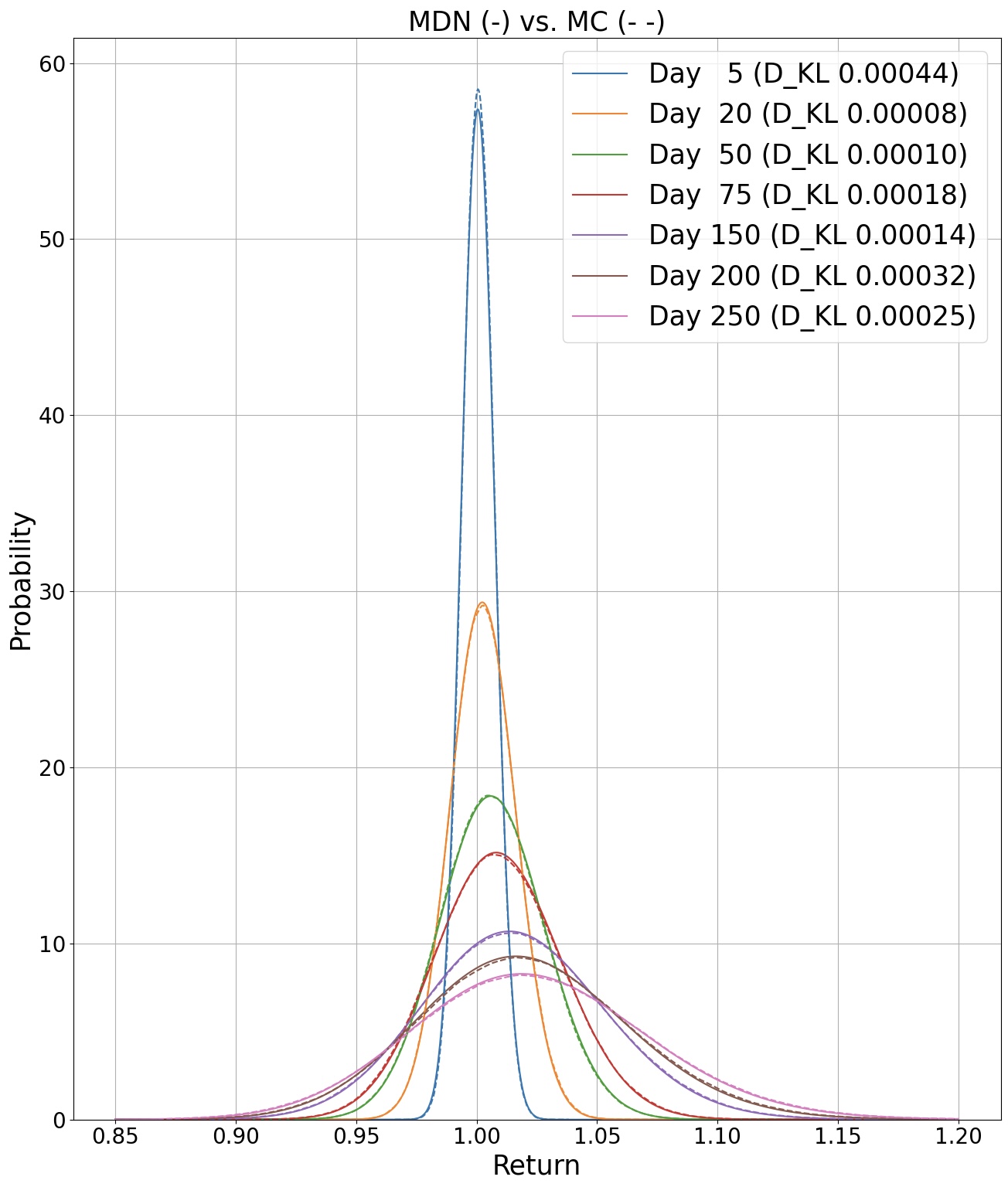}
    }\hfill
    \subfloat[$w_1=0.75, w_2=0.25$\label{fig:GBM_LV_corr_time_variate_1d}]{%
        \includegraphics[width=0.3\textwidth]{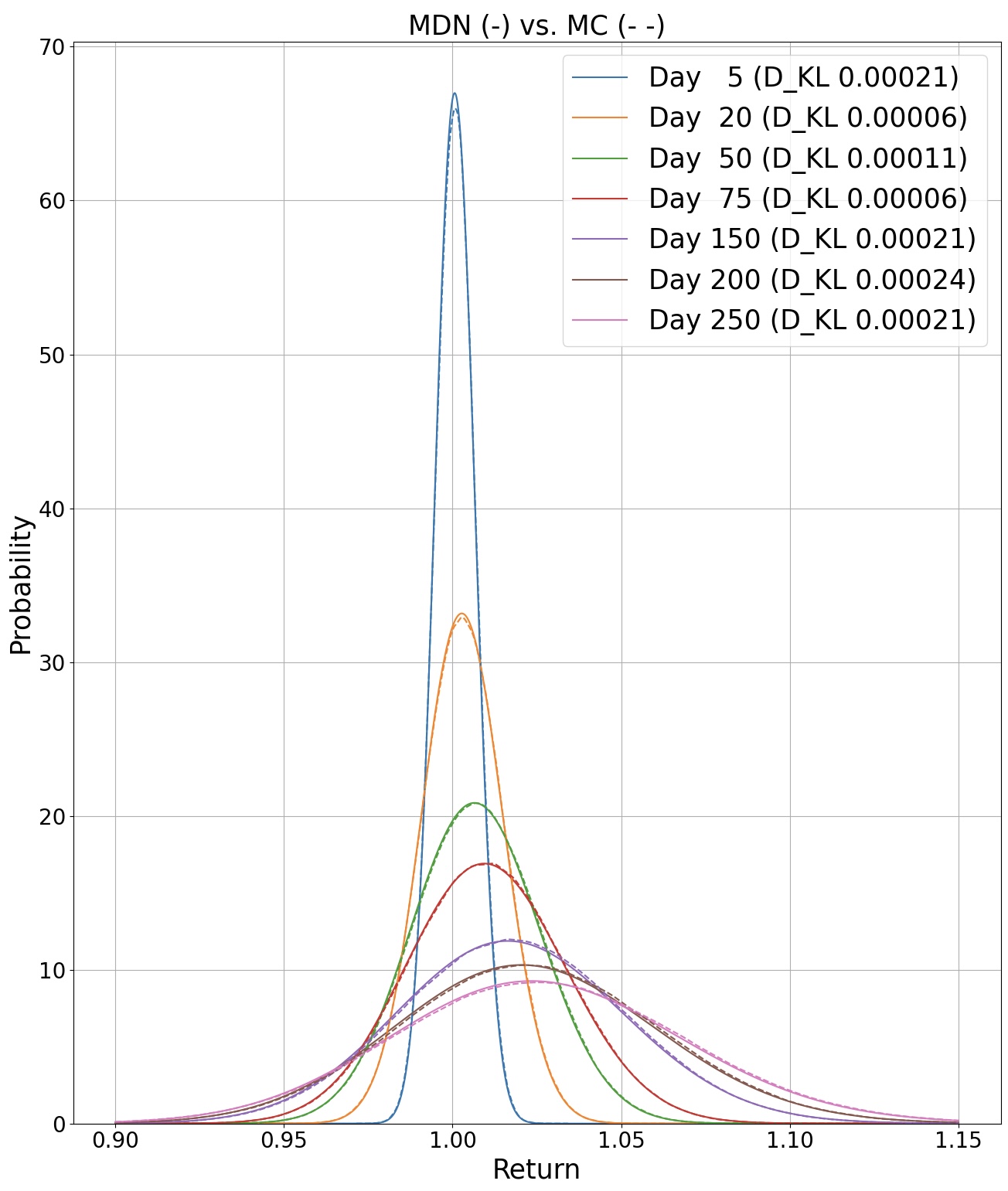}
    }

    \caption{(a) Time-varying interest rates and dividend yields. (b),(c)\&(d) MDN vs. MC distributions of weighted basket returns with KL divergence at different maturities for different weight configurations.}
    \label{fig:GBM_LV_corr_time_variate_1}
\end{figure}

\begin{figure}[hbtp]
    \centering
    \includegraphics[width=.5\textwidth]{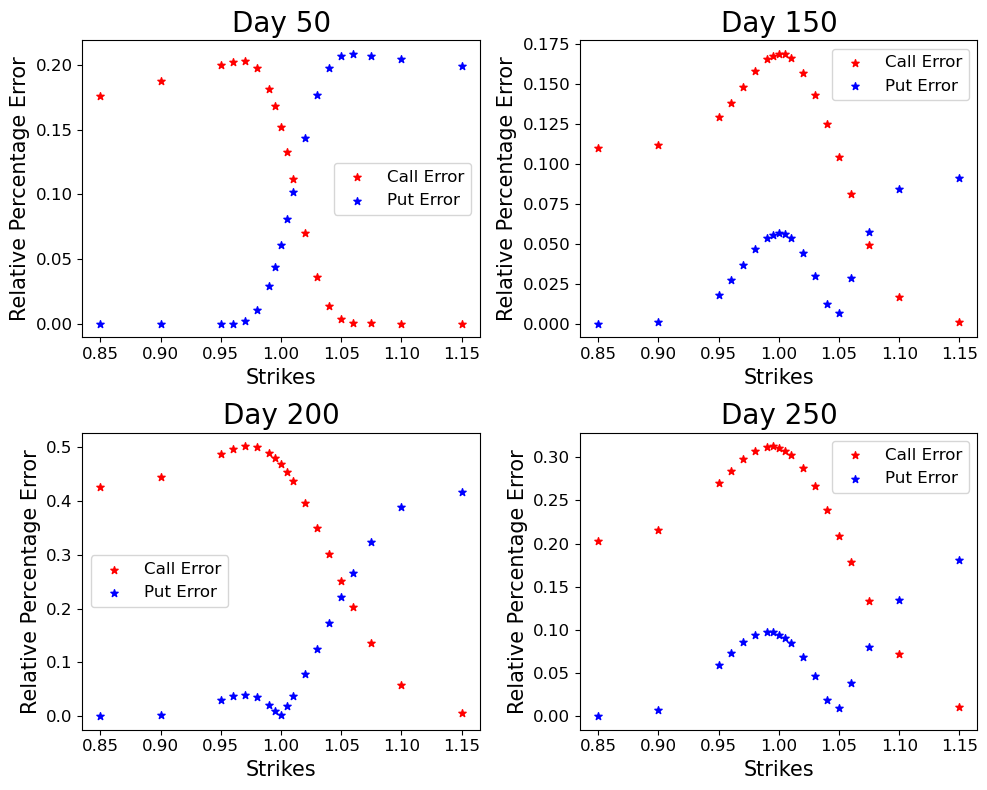}
    \caption{Relative percentage error in European call and put option prices based on MDN vs. MC pricing.}
    \label{fig:GBM_LV_N2CSign_price_error_02}
\end{figure}

Figure \ref{fig:GBM_LV_corr_time_variate_1a} shows the time series of the risk-free interest rate and the dividend yields for the two assets in the basket. Using this market condition, we analyze three different combinations of the basket weights. The resulting distributions of the basket’s weighted returns, estimated by the mixture density network (MDN), are compared with those obtained via Monte Carlo (MC) simulation in Figures \ref{fig:GBM_LV_corr_time_variate_1b}, \ref{fig:GBM_LV_corr_time_variate_1c} and \ref{fig:GBM_LV_corr_time_variate_1d}, each corresponding to different weight vectors. These comparisons are carried out across multiple maturities. For each case, the Kullback–Leibler (KL) divergence between the MDN and MC distributions is reported alongside the corresponding maturity, denoted by "D\_KL".

Under the same market conditions, and using the basket weight from Figure \ref{fig:GBM_LV_corr_time_variate_1c}, we present the relative percentage errors between the European option prices computed from the MDN-based distribution and MC-based distribution in Figure \ref{fig:GBM_LV_N2CSign_price_error_02}.\\

\FloatBarrier
\begin{figure}[hbtp]
    \centering
    % Top row: Single figure
    \subfloat[\label{fig:GBM_LV_corr_time_variate_2a}]{%
        \includegraphics[width=0.3\textwidth]{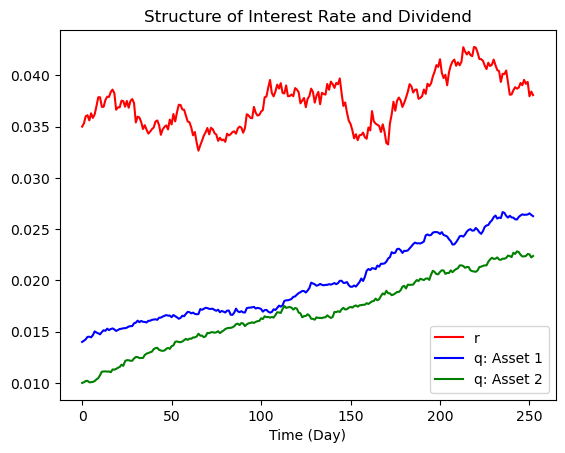}
    }

    \vspace{1ex} % vertical space between rows

    % Bottom row: Three side-by-side
    \subfloat[$w_1=0.25, w_2=0.75$\label{fig:GBM_LV_corr_time_variate_2b}]{%
        \includegraphics[width=0.3\textwidth]{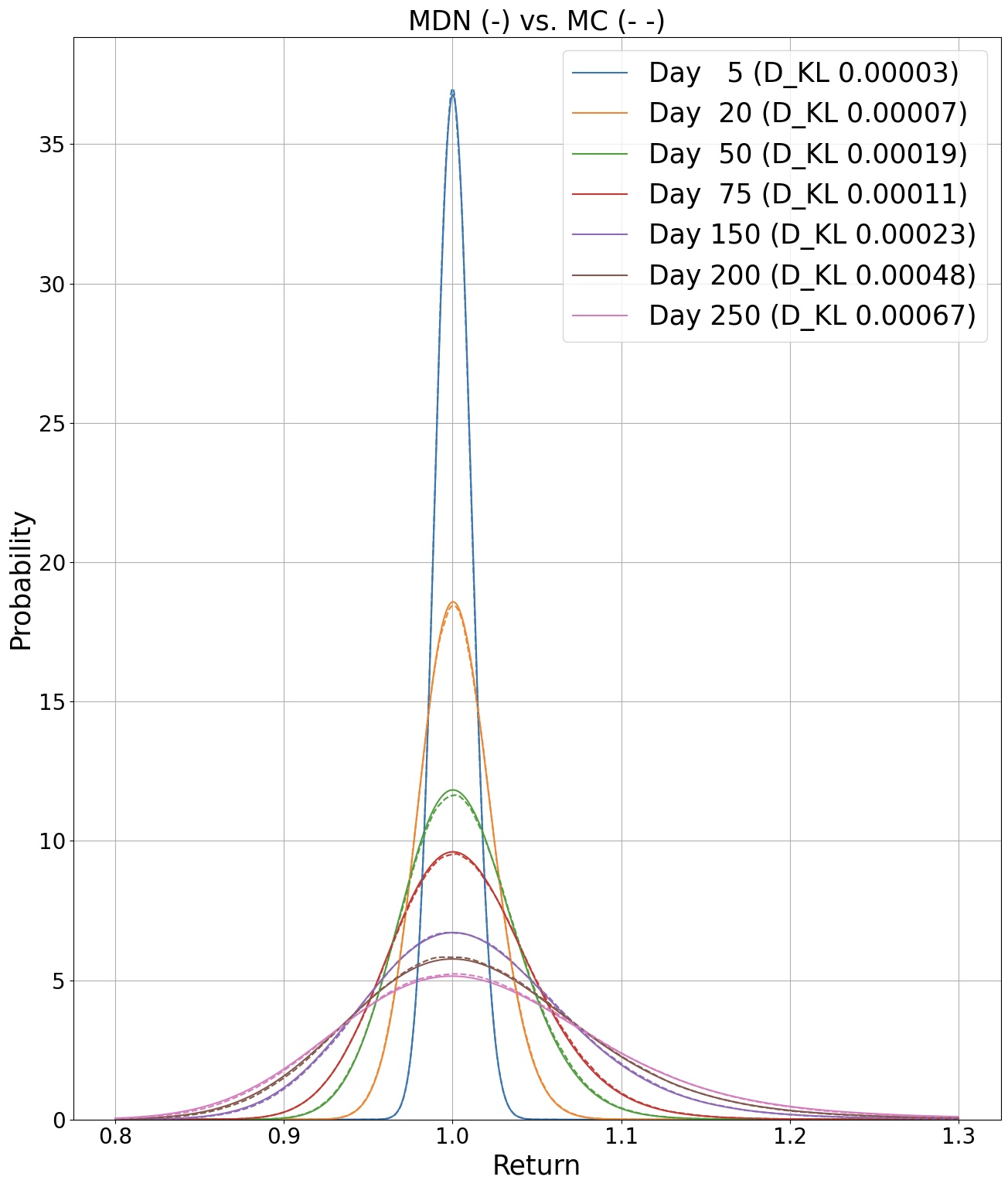}
    }\hfill
    \subfloat[$w_1=0.5, w_2=0.5$\label{fig:GBM_LV_corr_time_variate_2c}]{%
        \includegraphics[width=0.3\textwidth]{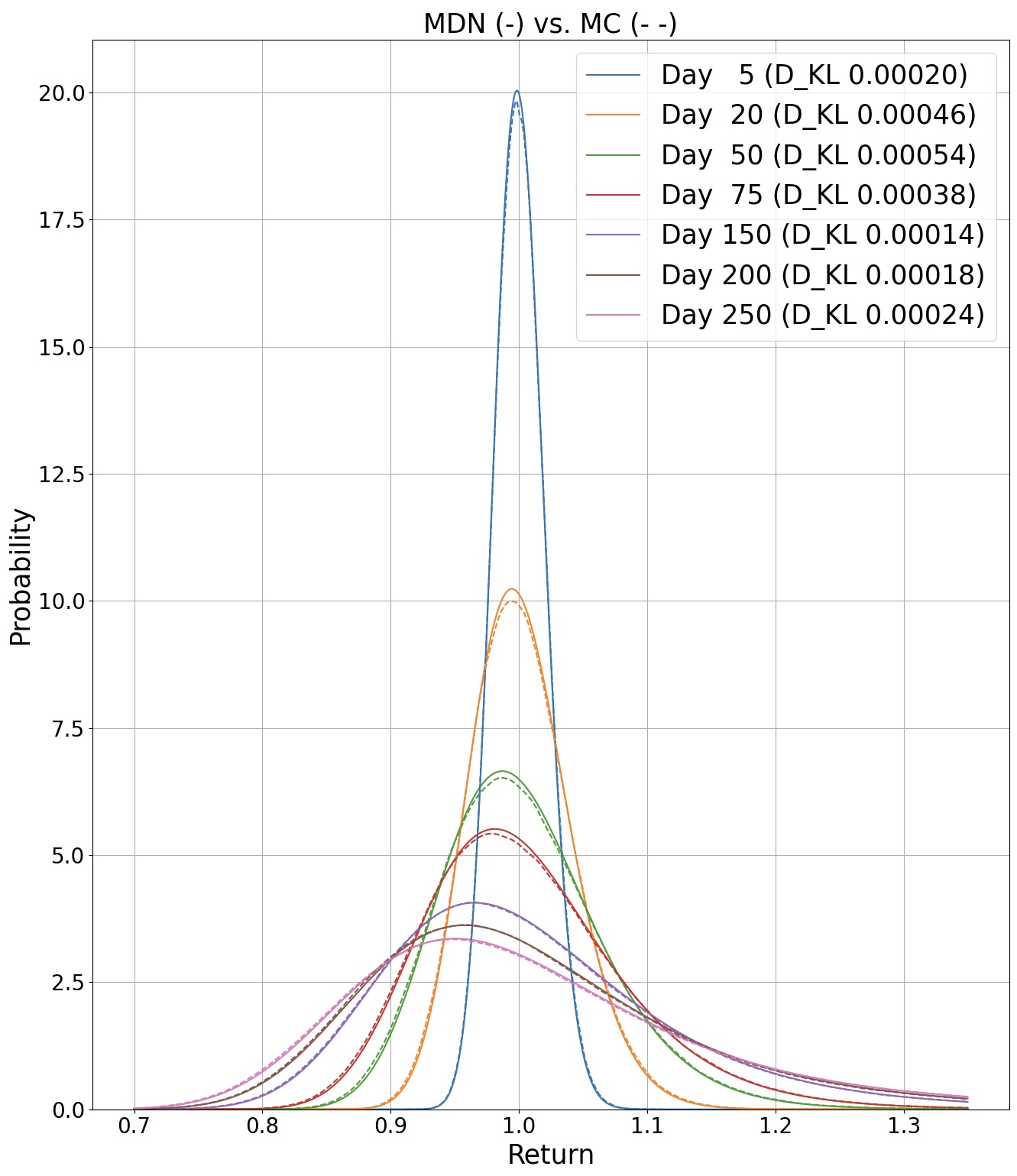}
    }\hfill
    \subfloat[$w_1=0.75, w_2=0.25$\label{fig:GBM_LV_corr_time_variate_2d}]{%
        \includegraphics[width=0.3\textwidth]{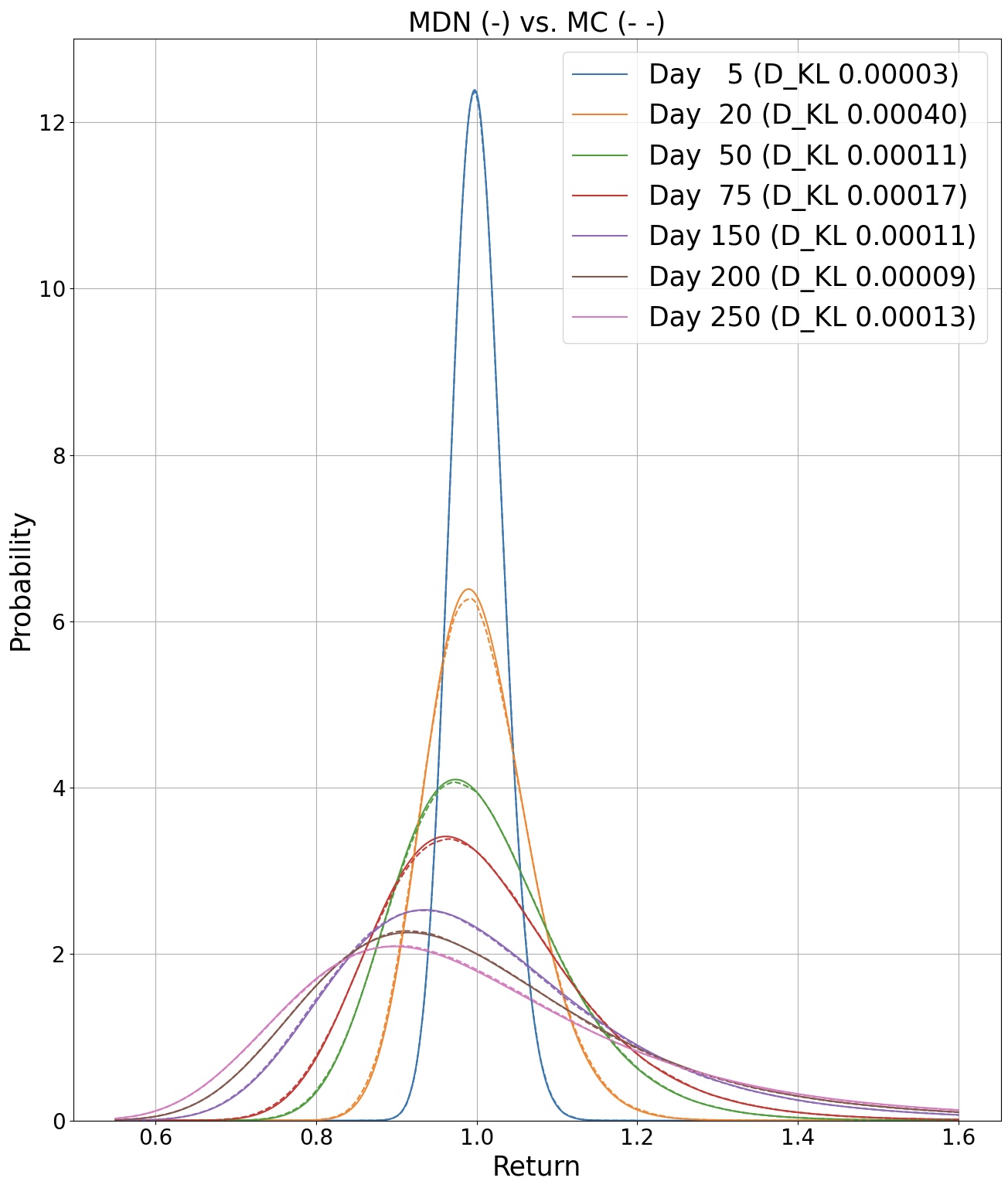}
    }

    \caption{(a) Time-varying interest rates and dividend yields. (b),(c)\&(d) MDN vs. MC distributions of weighted basket returns with KL divergence at different maturities for different weight configurations.}
    \label{fig:GBM_LV_corr_time_variate_2}
\end{figure}

\begin{figure}[hbtp]
    \centering
    \includegraphics[width=.5\textwidth]{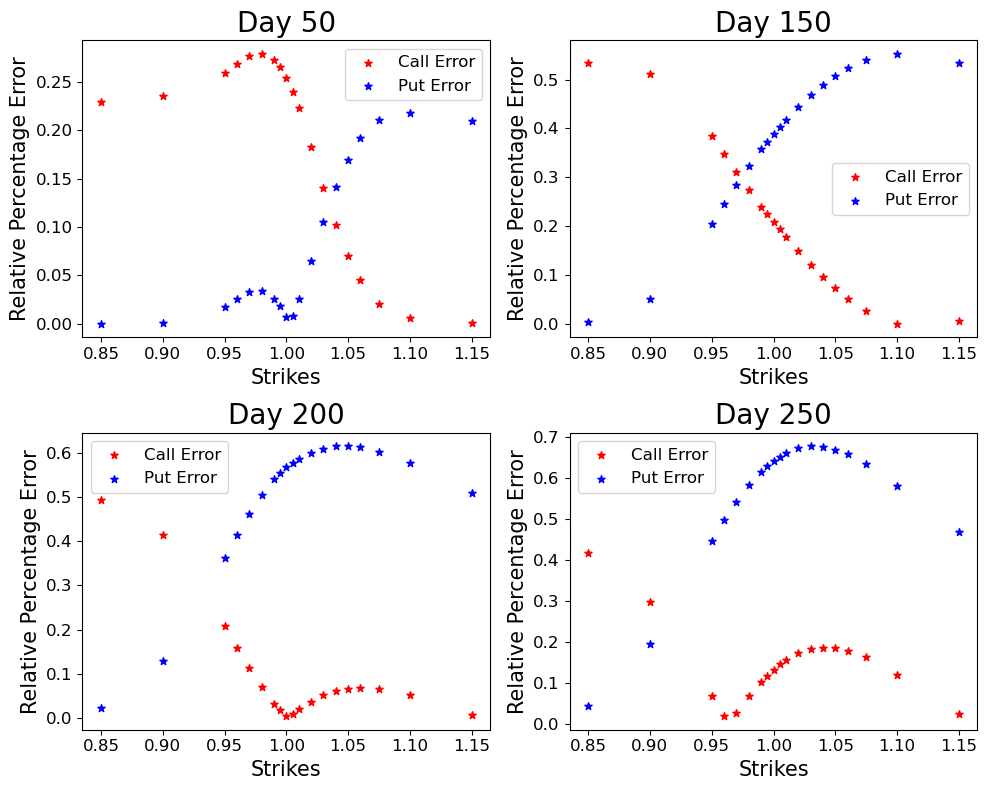}
    \caption{Relative percentage error in European call and put option prices based on MDN vs. MC pricing.}
    \label{fig:GBM_LV_N2CSign_price_error_11}
\end{figure}

\begin{figure}[hbtp]
    \centering
    \includegraphics[width=.5\textwidth]{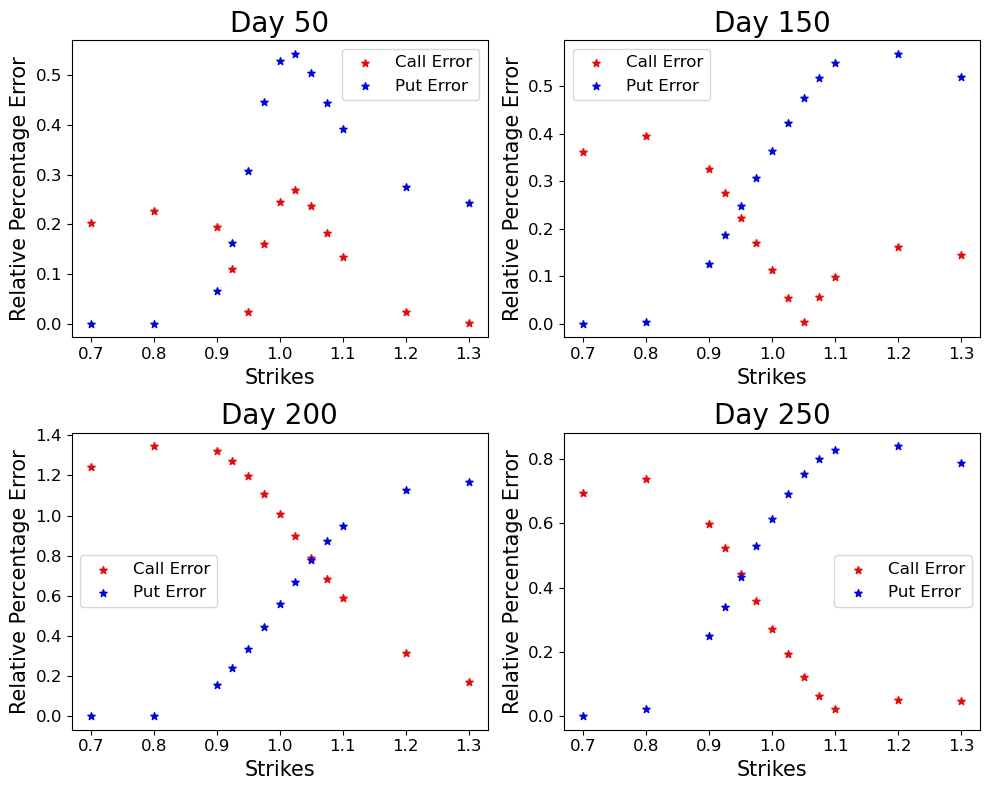}
    \caption{Relative percentage error in European call and put option prices based on MDN vs. MC pricing.}
    \label{fig:GBM_LV_N2CSign_price_error_12}
\end{figure}

\noindent In a second scenario, we modify the correlation matrix to
\[
\mathbf{R} = \begin{pmatrix}
1 & -0.4849 \\
-0.4849 & 1
\end{pmatrix},
\]
and update the local-volatility function parameters for the two assets to
$$a_{loc}=[0.6775,0.7475],\;b_{loc}=[0.3023,0.1543],\; c_{loc}=[0.3951,0.1214].$$

The risk-free interest rate and dividend yields for two assets are now given in Figure \ref{fig:GBM_LV_corr_time_variate_2a}. For three different sets of basket weights, Figures~\ref{fig:GBM_LV_corr_time_variate_2b}, \ref{fig:GBM_LV_corr_time_variate_2c}, and \ref{fig:GBM_LV_corr_time_variate_2d} present a comparison between the MDN-estimated and Monte Carlo-simulated distributions of the weighted basket return across various maturities. The associated Kullback–Leibler (KL) divergence between the two distributions is also reported for each case.

The resulting relative percentage errors in European option prices, corresponding to the basket weights used in Figure~\ref{fig:GBM_LV_corr_time_variate_2b} and \ref{fig:GBM_LV_corr_time_variate_2c}, are reported in Figure~\ref{fig:GBM_LV_N2CSign_price_error_11} and ~\ref{fig:GBM_LV_N2CSign_price_error_12} respectively.
\FloatBarrier
These numerical results demonstrate that the MDN can successfully learn the conditional distribution based on the parameters of the local volatility function. Notably, the MDN is also parametrized by the basket weights, providing additional flexibility: if the basket composition changes, the trained model remains applicable for pricing options on the new weighted return, eliminating the need to retrain a separate model.\\

\noindent \textbf{Reproducibility and Runtime: } 
All experiments were conducted on an Apple M1 Pro processor with 16 GB of RAM. Training on 20 million samples (batch size = 100,000) required approximately 6.5 minutes per epoch. The average inference latency per market configuration was about 3.4 milliseconds, demonstrating the model’s suitability for real-time or large-scale pricing applications.

\section{Conclusion}\label{sec:conclusion}

In this work, we proposed a deep learning-based framework for pricing European basket options under geometric Brownian motion (GBM) dynamics with time-varying market parameters. Our approach leverages a mixture density network (MDN) to approximate the conditional distribution of the weighted return of a basket of assets, taking as input the time series of risk-free interest rates, dividend yields, volatilities, and correlation structures.

We evaluated the method under two GBM settings—one with time-varying volatility and another with local volatility. In both cases, the MDN accurately recovered the conditional distribution across a broad range of input scenarios. Option prices computed using the learned distributions closely matched benchmark Monte Carlo (MC) estimates, with relative pricing errors typically within a few percentage points. These results demonstrate the MDN’s effectiveness as a surrogate distribution model for pricing.

A key component of our architecture is the use of truncated path signatures to represent the temporal evolution of model parameters. This feature representation enabled the MDN to generalize across diverse time-varying environments. Additionally, by incorporating basket weights as part of the input, the model can adapt to different portfolio compositions without retraining, offering practical flexibility.

Overall, the results suggest that MDNs, when paired with appropriate feature engineering and training design, provide an efficient and accurate alternative to traditional simulation-based pricing methods. Once trained, the MDN enables rapid inference, making it particularly suitable for real-time or high-frequency pricing applications.

Future research could extend this approach to modelling multivariate output distributions, more complex derivative products, or alternative market dynamics, including stochastic volatility, rough volatility, and jump processes. Additionally, the framework could be adapted with minimal changes to learn univariate distributions of other functionals of the asset paths, such as the maximum or minimum of a basket.

\newpage
%\nocitep{*}

\addcontentsline{toc}{section}{References}
\bibliography{biblist}
\bibliographystyle{alpha}

\end{document}